\documentclass{article}
\usepackage[margin=1in,paperheight=11in,paperwidth=8.5in]{geometry}
\usepackage{authblk}
\usepackage{babel}
\usepackage{mathtools}
\usepackage{amssymb} 
\usepackage{float}
\usepackage{enumitem}
\usepackage{amsthm}
\usepackage{amsmath}
\usepackage{amsfonts}
\usepackage{array}
\usepackage{amsopn}
\usepackage{natbib}
\usepackage{enumitem}
\DeclarePairedDelimiterX\Basics[1](){ #1}
\usepackage[utf8]{inputenc}
\usepackage{color}
\usepackage{subfigure}
\usepackage{enumitem}
\usepackage{float}
\usepackage{pifont}
\usepackage{algorithm, algpseudocode, algcompatible}
\usepackage[T1]{fontenc}
\usepackage{etoolbox}
\usepackage{float}
\usepackage{arydshln}
\usepackage{lscape}
\usepackage{bm}
\usepackage{multirow}
\usepackage{lineno}
\usepackage{diagbox}
\usepackage{array, booktabs, multirow}
\usepackage{graphicx}
\usepackage{caption}
\usepackage{bbm}
\usepackage{setspace}
\usepackage{comment}
\DeclareCaptionJustification{double}{\doublespacing}
\definecolor{aoe}{rgb}{0.0, 0.5, 0.0}
\definecolor{airforceblue}{rgb}{0.36, 0.54, 0.66}
\definecolor{camel}{rgb}{0.76, 0.6, 0.42}
\definecolor{armygreen}{rgb}{0.29, 0.33, 0.13}
\definecolor{carrotorange}{rgb}{0.93, 0.57, 0.13}
\definecolor{darkseagreen}{rgb}{0.56, 0.74, 0.56}

\usepackage{xcolor}

\newtheorem{definition}{Definition}[section]
\newtheorem{proposition}{Proposition}[section]
\newtheoremstyle{example}{}{}{}{}{\bfseries}{\smallskip}{\newline}{}
\theoremstyle{example}
\newtheorem{example}{Example}

\newcommand*\samethanks[1][\value{footnote}]{\footnotemark[#1]}

\newcounter{noteMCctr} 

\newcounter{noteCZctr} 

\newcounter{noteXPctr} 

\usepackage[colorlinks=true,linkcolor=red,filecolor=green,citecolor=blue]{hyperref}

\title{\textbf{Graph Neural Networks for Forecasting Multivariate Realized Volatility with Spillover Effects}}
\author[1,2,3]{Chao Zhang\thanks{The first two authors contributed equally to this work. Correspondence to: Chao Zhang <chao.zhang@stats.ox.ac.uk>. 
The authors thank the Oxford Suzhou Centre for Advanced Research for providing the computational facilities. An earlier version of this article circulated under the title ``Graph Neural Networks for Forecasting Realized Volatility with Nonlinear Spillover Effects''. First draft: January 2023.}} 
\author[3,4]{Xingyue Pu\samethanks}
\author[1,2,3,5]{Mihai Cucuringu}
\author[3,4]{Xiaowen Dong}

\affil[1]{\small Department of Statistics, University of Oxford, Oxford, UK}
\affil[2]{\small Mathematical Institute, University of Oxford, Oxford, UK}
\affil[3]{\small Oxford-Man Institute of Quantitative Finance, University of Oxford, Oxford, UK}
\affil[4]{\small Department of Engineering Science, University of Oxford, Oxford, UK}
\affil[5]{\small The Alan Turing Institute, London, UK}
\date{This version: May 2023}

\begin{document}

\maketitle
\begin{abstract}
We present a novel methodology for modeling and forecasting multivariate realized volatilities using customized graph neural networks to incorporate spillover effects across stocks. The proposed model offers the benefits of incorporating spillover effects from multi-hop neighbors, capturing nonlinear relationships, and flexible training with different loss functions. 
Our empirical findings provide compelling evidence that incorporating spillover effects from multi-hop neighbors alone does not yield a clear advantage in terms of predictive accuracy. However, modeling nonlinear spillover effects enhances the forecasting accuracy of realized volatilities, particularly for short-term horizons of up to one week. Moreover, our results consistently indicate that training with the Quasi-likelihood loss leads to substantial improvements in model performance compared to the commonly-used mean squared error. A comprehensive series of empirical evaluations in alternative settings confirm the robustness of our results.
% previous abstract
% We propose a novel methodology for modeling and forecasting multivariate realized volatilities using graph neural networks. This approach extends the work of \cite{GHAR_zhang2022graph} and explicitly incorporates the spillover effects from multi-hop neighbors and nonlinear relationships into the volatility forecasts. Our findings provide strong evidence that the information from multi-hop neighbors does not offer a clear advantage in terms of predictive accuracy. However, modeling the nonlinear spillover effects significantly enhances the forecasting accuracy of realized volatilities over up to one month. Our model is flexible and allows for training with different loss functions, and the results generally suggest that using Quasi-likelihood as the training loss can significantly improve the model performance, compared to the commonly-used mean squared error. A comprehensive series of evaluation tests and alternative model specifications confirm the robustness of our results.
\end{abstract}

\bigskip


% \newpage
% \tableofcontents
% \newpage

%%%%%%%%%%%%%%%%%%%%%%%%%%%%%%%%%%%%%%%%%%%%%%%%%%%%%%%%%%%%%%
%%%%%%%%%%%%%%%%%%%%%% Introduction %%%%%%%%%%%%%%%%%%%%%%%%%%
%%%%%%%%%%%%%%%%%%%%%%%%%%%%%%%%%%%%%%%%%%%%%%%%%%%%%%%%%%%%%%

\section{Introduction}
Modeling and forecasting stock return volatility plays a crucial role in the theory and practice of finance. Extensive attention has been devoted to this subject within the literature, encompassing numerous ARCH, GARCH, and stochastic volatility models. Due to the availability of high-frequency data, realized volatility (RV), calculated from the sum of squared intraday returns, has gained popularity in recent years. For example, \cite{corsi2009simple} put forward the Heterogeneous Autoregressive (HAR) for predicting daily RVs using various lagged RV components over different time horizons. While {these methods} provided valuable insights into the dynamic dependence of volatilities, they neglected the volatility spillover effect among assets, as highlighted by \cite{bollerslev2018risk}.

The \textit{volatility spillover effect} is referred to as the phenomenon that some  big shocks of a specific asset (or market) may have an influence on the volatilities of other assets (or markets). Therefore, the discovery of volatility spillover effects is expected to benefit the understanding and forecasting of volatilities. For example, \cite{buncic2016global} documented that the VIX of the U.S. market plays an important role in forecasting the volatilities of other global assets markets. \cite{degiannakis2017forecasting} examined the cross-asset spillover effects from stocks, currencies, and commodities to improve the prediction of RV of crude oil. \cite{bollerslev2018risk, li2021forecasting} utilized the commonality in risk structures to improve the forecasting of future volatility. 
% \cite{GHAR_zhang2022graph} employed various financial networks to harness the spillover effect and enhance the accuracy of individual volatility forecasts. \par

There is a number of studies dedicated to incorporating the spillover effect into volatility modeling, e.g. BEKK-GARCH (\cite{engle1995multivariate}) and VAR-GARCH (\cite{ling2003asymptotic}).  In terms of modeling RV, \cite{wilms2021multivariate} employed Vector Autoregression (VAR) to obtain the multivariate volatility forecasts for stock market indices. However, in high-dimensional scenarios, the aforementioned models may deliver poor out-of-sample forecasts due to the curse of dimensionality, as emphasized by \cite{callot2017modeling}. Most recently, \cite{GHAR_zhang2022graph} introduced graph-based methods to capture the volatility spillover effects, and proposed a parsimonious model to augment HAR via neighborhood aggregation on a graph that represents a financial network, denoted as Graph HAR (or GHAR). In these graphs, each asset is modeled as a node and an edge connecting two nodes encodes the existence of the spillover effect between their volatilities. GHAR leverages neighborhood aggregation to generate a new covariate over the graph for each underlying asset and enhance the accuracy of individual volatility forecasts.

\begin{figure}[H]
    \centering
    \caption{An illustration of multi-hop and nonlinear volatility spillover.}
    \includegraphics[width=0.8\textwidth]{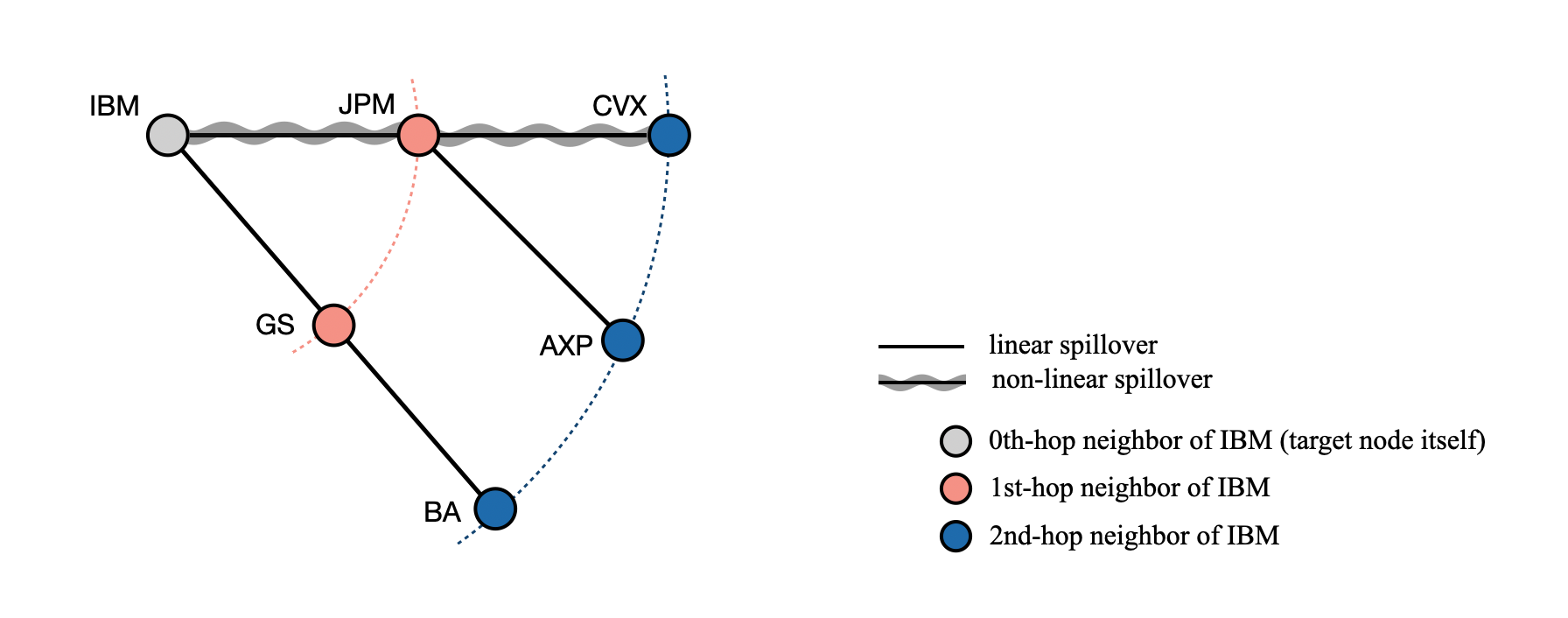}
    \caption*{\textit{Note:} The target node represents the volatility of IBM. The connections are only for illustration, and hence not necessarily consistent with our experiments.}
    \label{fig:illustration}
\end{figure}

One natural question following GHAR is whether there exists any spillover effect between nodes that is beyond one step, a.k.a. \textit{multi-hop neighbors} (see detailed definitions in Section \ref{sec:pre_graph}). For example, as illustrated in Figure \ref{fig:illustration}, for the target node (i.e. IBM), in addition to the spillover effect of 1st-hop neighbors (i.e. JPM and GS), we are also interested in whether there is any spillover effect from 2nd-hop neighbors (i.e. AXP, CVX, and BA). 
% Similarly, taking into account any potential spillover effect from multi-hop neighbors would lead to improved volatility forecasting. 
To the best of our knowledge, spillover effects from multi-hop neighbors have not yet received much attention in the volatility modeling literature.\footnote{The 2nd-hop connections have been studied in the context of cascading effects of financial networks, e.g. \cite{acemoglu2010cascades}, where the shocks that occur to an individual firm would propagate through the rest of the economy. Consequently, the downstream firms more than one hop away may also suffer from the impact.} 

In addition to multi-hop effects, another interesting question is whether the volatility spillover is \textit{nonlinear}. Most of the previous studies focus on the linear relationships across assets or markets, such as \cite{degiannakis2017forecasting, wilms2021multivariate, GHAR_zhang2022graph}. In a few scenarios, the existence of nonlinear volatility spillover effects has also been discovered and examined.  For example, \cite{choudhry2016stock} documented the existence of significant nonlinear spillover effects among four major markets, i.e. the U.S., Canada, Japan, and the U.K., via a nonlinear causality test proposed by \cite{bai2010multivariate}. \cite{wang2018oil} attempted to capture the nonlinear relationship between the volatilities of stocks and crude oil, by incorporating the asymmetric effect of oil prices and regime shifts. 
% However, their findings revealed limited evidence supporting the superiority of nonlinear models over linear alternatives in forecasting the volatility of stocks in relation to commodities.

While the incorporation of multi-hop neighbors expands the set of features and the potential presence of nonlinear spillover effects introduces new functional forms to describe volatility dynamics, it is also worth emphasizing that the choice of \textit{estimation criterion (EC)}, representing the objective function for estimating model parameters, plays a crucial role. This follows the perspective that a statistical forecasting model typically consists of three essential components: (i) feature set, (ii) model specification, and (iii) EC. 
%Viewed from this perspective, the incorporation of multi-hop effects expands the set of features, while the potential presence of nonlinear spillover effects introduces new functional forms to describe volatility dynamics. It is also worth emphasizing that the choice of EC, representing the objective function for estimating model parameters, plays a crucial role. 
Traditional econometric models, such as GARCH, commonly employ conditional quasi-likelihood (QLIKE) based on normal distributions for parameter estimation. Conversely, models focused on forecasting realized volatilities, such as HAR, often utilize the mean squared error (MSE) as their EC. Therefore, an important question arises as to whether a preferred EC exists\footnote{\cite{cipollini2020realized} conducted an empirical evaluation of the influence of various EC on linear models and found that using the QLIKE yields slightly better forecasts.}, especially when combined with the aforementioned aspects, namely the effect of multi-hop neighbors and non-linear relationships.

In the present work, we explore these three questions using graph neural networks (GNNs). GNNs are a class of deep learning models designed for performing inference on graphs and graph-structured data. 
% They are capable of learning vector representations of nodes or the graph structure as a whole. The node representations, for example, can be further applied to node classification or regression. 
They are capable of learning node and graph-level representations that are useful for a wide range of tasks involving graph analysis, such as node classification, node regression, and graph clustering.
% Various GNN variants employ the concept of neighborhood aggregation to formulate layer-wise forward propagation rules. For instance, GNN layers often calculate the mean of the representations of connected nodes, followed by a nonlinear activation function. 
GNNs have demonstrated successful applications in various financial domains, including stock movement prediction (\cite{gnn_stock_chen2018incorporating, gnn_stock_sawhney2020deep}), credit risk prediction (\cite{gnn_finance_risk_wang2019semi, gnn_finance_risk_liang2021credit}) and payment fraud detection (\cite{gnn_fraud_liu2018heterogeneous, gnn_fraud_liu2019geniepath}).

In particular, we design a GNN-based framework that considers the topological characteristics of volatility spillover effects.\footnote{A contemporaneous study by \cite{chen2022multivariate} employed GNN for intraday volatility forecasting, but their method faced limitations in terms of interpretability and benchmarking challenges.} By replacing the linear neighborhood aggregation in the GHAR of \cite{GHAR_zhang2022graph} with a nonlinear operation, the proposed model is able to automatically learn the nonlinear spillover effects. Furthermore, the multi-layer setting of GNNs allows us to explore this nonlinearity in the multi-hop setting, i.e. spillover to neighbors that are more than one hop away in the financial network. Finally, an inherent advantage of our model lies in its flexibility to accommodate various EC during the training phase.

The main contributions of our work are summarized as follows. First, we examine the spillover effect from multi-hop neighbors in the financial graph, and observe that {the multi-hop spillover effect is not necessary}, as long as 0-hop and 1-hop are included. Second, we establish that the proposed {GNN model with nonlinear operations significantly improves the forecasting performance of GHAR}, indicating the existence of nonlinear spillover effects on 1-hop neighbors. Third, compared to MSE-trained models, models employing QLIKE as the EC generally achieve substantial improvements in predictive accuracy, highlighting the effectiveness of QLIKE in modeling the volatility process. Our proposed GNN model trained with QLIKE exhibits an average forecast error in MSE (resp. QLIKE) approximately 13\% (resp. 4\%) lower than that of the standard HAR model. Furthermore, we examine the robustness of our proposed models across various market conditions, an alternative data-splitting scheme, and an alternative universe, consistently observing enhanced prediction accuracy across all experimental settings.

The remainder of this paper is organized as follows. Section \ref{sec:gnn_overviews} contains preliminaries on the mathematical definitions of graphs, a brief review of GNN models, and two baseline models (HAR, GHAR). In Section \ref{sec:proposed_models}, we introduce the proposed model (GNNHAR), evaluation criterion, and forecast evaluation approaches. Section \ref{sec:experiment} outlines the experimental setup and provides the key out-of-sample results across various forecast horizons and market regimes. Furthermore, in Section \ref{sec:discuss}, we conduct an extensive analysis concerning the impact of QLIKE, nonlinearity, and multi-hop neighbors. In Section \ref{sec:robustness}, we perform several robustness tests. We conclude our work and highlight future research  directions in Section \ref{sec:conclusion}.

%%%%%%%%%%%%%%%%%%%%%%%%%%%%%%%%%%%%%%%%%%%%%%%%%%%%%%%%%%%%%%
%%%%%%%%%%%%%%%%%%%%%%     GNN      %%%%%%%%%%%%%%%%%%%%%%%%%%
%%%%%%%%%%%%%%%%%%%%%%%%%%%%%%%%%%%%%%%%%%%%%%%%%%%%%%%%%%%%%%

\section{Preliminaries}
\label{sec:gnn_overviews}

In this section, we summarize the preliminary concepts and models. In particular, we provide the mathematical definitions of graphs and multi-hop neighbors in Section \ref{sec:pre_graph}. In Section \ref{sec:gnn_models}, we briefly review two popular graph neural networks, that inspired our work. Section \ref{sec:preliminary_rv} revisits the baseline model HAR for forecasting realized volatilities, while Section \ref{sec:preliminary_ghar} reviews another baseline model GHAR.

\subsection{Graph definitions}\label{sec:pre_graph}

\begin{definition}[Graphs] \label{defn:graphs}
A graph $\mathcal{G}$ is defined as $\mathcal{G}= \{\mathcal{V}, \mathcal{E}\}$, where $\mathcal{V}=\left\{v_1, \ldots, v_N\right\}$ is a set of $N$ nodes and $\mathcal{E}$ is a set of edges, where $e_{i j}=\left(v_i, v_j\right) \in \mathcal{E}$ denotes an edge connecting node $v_i$ and node $v_j$.
\end{definition}

\begin{definition}[Adjacency matrix]
An adjacency matrix $\mathbf{A}$ is a square matrix whose dimension is $N \times N$, where $\mathbf{A}[i,j]$ represents the connection between $v_i$ and $v_j$ in the  graph $\mathcal{G}$. An adjacency matrix can be weighted, where $\mathbf{A}[i,j] \geq 0, \,\, \forall i,j$ represents the strength/intensity of the connection between nodes  $v_i$ and $v_j$. If $\mathbf{A}[i,j] \in \{0, 1\}, \,\,  \forall i,j$, the graph is a binary graph. The diagonal elements of $\mathbf{A}$ are all 0 since edges from a node to itself are typically not considered in graphs. In this article, we mainly consider \textbf{binary} graphs without self-connections. \end{definition}

\begin{definition}[$K$-hop neighbors] Following \cite{gnn_hop_feng2022powerful}, we use the $K$-hop neighbors of node $v$ to represent all the neighbors that have distance from node $v$ less than or equal to $K$, based on the shortest path distance (SPD) kernel. In contrast, $k$th-hop neighbors represent the neighbors with exact distance $k$ from node $v$. Finally, we denote $Q_{v, \mathcal{G}}^K$ as the set of $K$-hop neighbors of node $v$ in graph $\mathcal{G}$. 
\end{definition}

\begin{figure}[H]
    \centering
    \caption{Illustration of a graph and its corresponding adjacency matrices with multi-hop neighbors.}
    \includegraphics[width=0.99\textwidth]{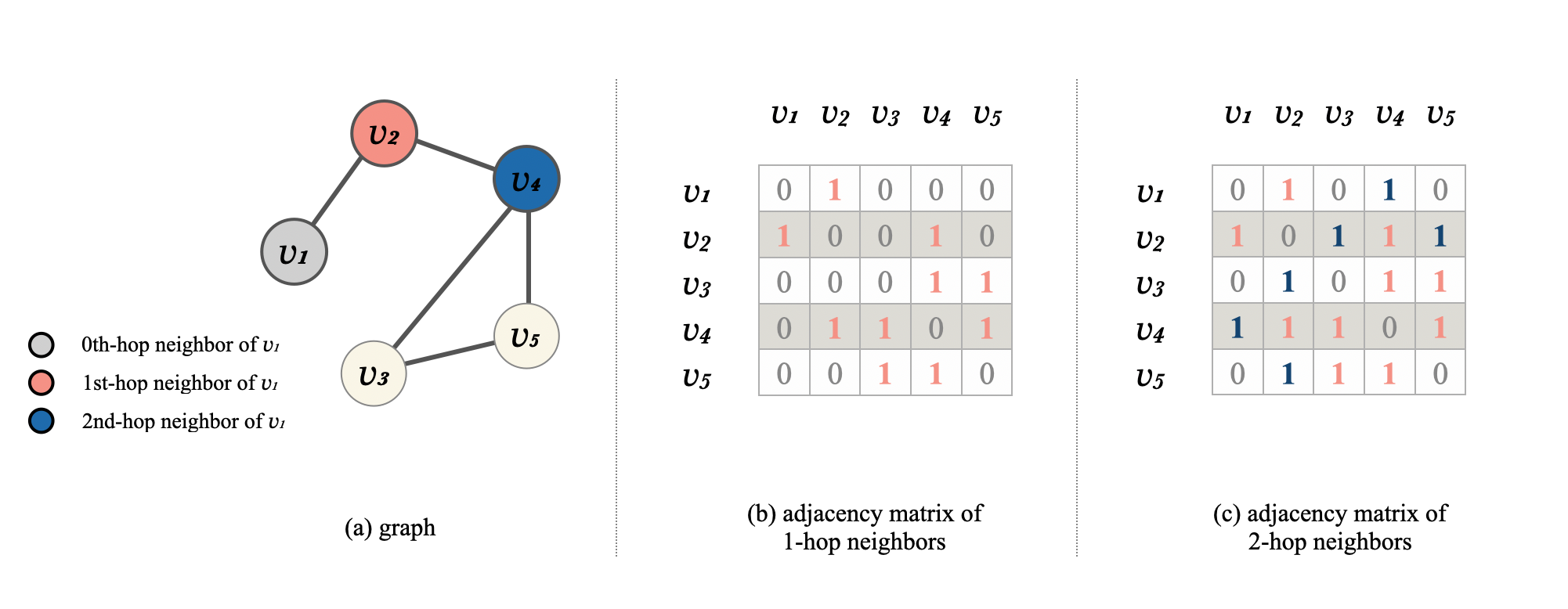}
    \label{fig:illustration_graphs_adj}
\end{figure}

\begin{example}[A graph with 5 nodes]\label{ex:adj_hop}
In Figure \ref{fig:illustration_graphs_adj}(a), we plot an example graph with 5 nodes and 5 undirected edges, where the node $v_1$ is colored as a target node. Nodes $v_2$ and $v_4$ are the 1st-hop and 2nd-hop neighbors of $v_1$,  respectively. Figure \ref{fig:illustration_graphs_adj}(b) shows its adjacency matrix.  Figure \ref{fig:illustration_graphs_adj}(c) is the adjacency matrix of the graph in (a) when considering 2-hop neighbors, where we write $Q_{v_1, \mathcal{G}}^1 = \{v_1, v_2\}$ and $Q_{v_1, \mathcal{G}}^2 = \{v_1, v_2, v_4\}$.
\end{example}

\subsection{A brief review on GNNs} \label{sec:gnn_models}

Graph neural networks (GNNs) are a class of deep learning models designed for performing inference on graphs. The main idea is to learn a vector representation for every node defined on a graph, while preserving both graph topology structure and node content information (\cite{gnn_wu2020survey}). The node representations, for example, can be further applied to node classification or regression. To this end, many GNN variants utilize the idea of \textbf{neighborhood aggregation} in developing the layerwise forward propagation rules. In essence, neighborhood aggregation effectively generates a node $v$'s representation by aggregating its own feature vector $\boldsymbol{h}_v \in \mathbb{R}^{D}$ and the feature vectors of its connected nodes $\boldsymbol{h}_u \in \mathbb{R}^{D}$, where $u \in Q_{v, \mathcal{G}}^1$. Common examples of aggregation functions include sum, mean, and maximum. Early attempts of GNNs, see \cite{gnn_early_scarselli2008graph} and \cite{gnn_early_dai2018learning}, update node representations by aggregating neighborhood information recursively until a stable equilibrium is reached.  
More efficiently, a novel notion of convolution operator can be defined on irregular graphs to process neighborhood aggregation in parallel, so-called graph convolution.\footnote{Convolution operation has been widely applied to regular grid data, e.g. image pixels. Recently, it has been extended to graph-structured data. More details can be found in \cite{gnn_survey_shuman2013emerging}.} A considerable number of GNN variants and architectures  are built from different graph convolution operators. We provide a brief introduction to a specific GNN architecture that is relevant to our volatility forecasting models.

\paragraph{Graph Convolutional Network} (GCN) was introduced by \cite{gcn_Kipf2017GCN}. It approximates the graph convolution with the following layer-wise propagation rule\footnote{The GCN propagation rule approximates the graph convolution with the first-order Chebyshev spectral polynomials (ChebyNet). It alleviates the gradient vanishing/exploding and stabilizes the training in ChebyNet by introducing a normalization step on $\boldsymbol{A}$.
More details about ChebyNet can be found in \cite{gcn_defferrard2016GCN}.}
\begin{equation}
\label{eq:gnn_literautre_gcn}
\vspace{-2mm}
\boldsymbol{H}^{(l+1)}= \sigma\left(\tilde{\boldsymbol{O}}^{-\frac{1}{2}} \tilde{\boldsymbol{A}} \tilde{\boldsymbol{O}}^{-\frac{1}{2}} \boldsymbol{H}^{(l)} \boldsymbol{\Theta}^{(l)} \right),
\vspace{-2mm}
\end{equation}%
where $\tilde{\boldsymbol{A}}=\boldsymbol{A}+\boldsymbol{I}_N$ is the adjacency matrix of the graph $\mathcal{G}$ with added self-connections, and $\tilde{\boldsymbol{O}}$ is a diagonal matrix with $\tilde{\boldsymbol{O}}_{i i}=\sum_j \tilde{\boldsymbol{A}}_{i j}$. $\tilde{\boldsymbol{O}}^{-\frac{1}{2}} \tilde{\boldsymbol{A}} \tilde{\boldsymbol{O}}^{-\frac{1}{2}}$ is the normalized adjacency matrix, introduced to stabilize the training of the GNN models. $\boldsymbol{\Theta}^{(l)} \in \mathbb{R}^{D^{(l)} \times D^{(l+1)}}$ is the layer-specific trainable weight matrix. $\boldsymbol{H}^{(l)} \in \mathbb{R}^{N \times D^{(l)}}$ is the matrix of node representations at $l$-th layer. $\boldsymbol{H}^{(0)}$ is the input node features. $\sigma(\cdot)$ denotes a nonlinear activation function, such as ReLU$(\cdot) = \operatorname{max}(0, \cdot)$.

When addressing various research problems, the above GNN layers can be combined with other deep learning layers in an end-to-end learning framework. Additionally, the exploration of multi-hop effects can be achieved by straightforwardly stacking multiple GNN layers within a model. A model that incorporates $K$-layer GNN layers is commonly referred to as a $K$-layer GNN model.

\begin{definition}[Receptive field] In a GNN model, the receptive field of a target node is the set of nodes of the graph that determine its representations; see \cite{gnn_hop_feng2022powerful, gnn_receptive_field_alon2020bottleneck}.
\end{definition}

\begin{proposition}
\label{proposition:GNN-k}
%After $K$ layers of graph convolution in a GNN model, every node has a receptive field of size $K$; see \cite{gnn_hop_feng2022powerful}.
After $K$ layers of graph convolution in a GNN model, every node representation is determined by the information from the nodes within $K$ hops; see \cite{gnn_hop_feng2022powerful}.
\end{proposition}
The above proposition states that the size of receptive field of every node is associated with the number of layers in a GNN model. It is found that \cite{gnn_receptive_field_alon2020bottleneck} when $K$ is unnecessarily large, any two nodes could easily have highly overlapping receptive fields, and consequently attain highly similar node representations, which leads to the problem of over-smoothing (see \cite{gnn_smoothness_li2018deeper, gnn_smoothness_chen2020measuring}). Therefore, a large $K$ does not always help, but on the contrary, it may lead to indistinguishable node representations, and thus weaken the forecasting or classification accuracy. 

\subsection{Forecasting RV with HAR} \label{sec:preliminary_rv}
% Assuming the price process $P_{i, t}$ 
% of a financial asset $i$ follows 
% \begin{equation}\label{eq:return}
%     \dd \log P_{i, t} = \mu_i \dd t + \sigma_{i,t} \dd W_t,
% \end{equation}
% where $\mu_i$ is the drift, $\sigma_{i,t}$ is the instantaneous volatility, and $W_t$ is the standard Brownian motion. The integrated variance (IV) of asset $i$ at day $t$ is defined as 
% \begin{equation}
%     \textsc{IV}_{i, t} = \int_{t-1}^t \sigma^2_{i, s} \dd s.
% \end{equation}% \par

Let $P_{i, t}$ denote the price of asset $i$ and $r_{i,t}=\log \left(P_{i, t} / P_{i, t-1}\right)$ be its log-return at day $t$. The standard approach for modeling return data is to use the decomposition
% \vspace{-5mm}
\begin{equation}
r_{i, t}=\mu_{i, t}+X_{i, t},
\end{equation}
where $\mu_{i, t}$ denotes the (conditional) mean of the return, $X_{i, t}$ is a diffusion term which may be modeled as
\begin{equation}\label{eq:vol_decomp}
X_{i, t}=\sigma_{i, t} \varepsilon_{i, t}, \quad \varepsilon_{i, t} \sim \operatorname{IID}(0,1),
\end{equation}
where $\sigma_{i, t}$ is often referred to as the volatility function, and $\varepsilon_{i, t}$ is assumed to be independent of $\sigma_t$. 

\cite{andersen2001distribution, barndorff2002econometric} showed that the sum of squared intraday returns is a consistent estimator of the unobserved $\sigma_{i, t}^2$. Therefore, the daily RV for a particular asset $i$ at day $t$ is defined as 
\begin{equation}
    {RV}_{i, t} =  \sum_{l=1}^{M} r_{i, t(l)}^{2}, 
\end{equation}
where $r_{i, t(l)}$ is the $l$-th $\Delta$-min log returns during day $t$, i.e.  $r_{t(l)} = \log p_{t(l\Delta)} - \log p_{t((l-1)\Delta)}$, and $p_{t(l\Delta)}$ is the price at time $l\Delta$ at day $t$. We refer to $\boldsymbol{RV}_{t} = ({RV}_{1, t}, \dots, {RV}_{N, t})^{\prime}$ as the vector of cross-sectional realized volatilities. In this article, we consider $5$-min windows in a trading day, following \cite{liu2015does}.\footnote{
We also adopt the subsampling averaging method (see \cite{sheppard2010financial, andersen2011realized, varneskov2013role}) to improve the above $RV$ estimation, which uses all $\Delta$-minute returns, not just non-overlapping ones.}

\cite{corsi2009simple} proposed a Heterogeneous Autoregressive Regression (HAR) model for modeling and forecasting the RV where the lagged daily, weekly, and monthly volatility components are incorporated as features. For a given asset $i$, its RV of day $t$ is modeled as 
\vspace{-2mm}
\begin{equation}
\label{eq:har}
\textbf{HAR}: {RV}_{i, t} = \alpha + \beta_{d} {{RV}}_{i, t-1}+\beta_{w} {{RV}}_{i, t-5:t-2}+ \beta_{m} {{RV}}_{i, t-22:t-6}+u_{i, t},
\vspace{-2mm}
\end{equation}
where ${{RV}}_{i, t-5:t-2} = \frac{1}{4} \sum_{k=2}^{5} {RV}_{i, t-k}, {{RV}}_{i, t-22:t-6} = \frac{1}{17} \sum_{k=6}^{22} {RV}_{i, t-k}$ denote the weekly and monthly lagged RV, respectively. The choice of a daily, weekly, and monthly lag is aiming to capture the long-memory dynamic dependencies observed in most RV series.

\subsection{Graph HAR (GHAR)} \label{sec:preliminary_ghar}
\cite{GHAR_zhang2022graph} augmented the HAR model to capture the volatility spillover effect via neighborhood aggregation on graphs. Denote $\boldsymbol{V}_{:t-1} = \left[\boldsymbol{RV}_{t-1}, \boldsymbol{RV}_{t-5:t-2}, \boldsymbol{RV}_{t-22:t-6} \right] \in \mathbb{R}^{N \times 3}$, GHAR is defined as
\begin{equation}\label{eq:ghar_1hop}
\vspace{-2mm}
\begin{aligned}
\textbf{GHAR}(\boldsymbol{A}): \quad \boldsymbol{RV}_{t} = \boldsymbol{\alpha} &+ \beta_d \boldsymbol{RV}_{t-1} + \beta_w \boldsymbol{RV}_{t-5:t-2} + \beta_m \boldsymbol{RV}_{t-22:t-6}  \\
    &+ \gamma_d \boldsymbol{W} \cdot \boldsymbol{RV}_{t-1} + \gamma_w \boldsymbol{W} \cdot \boldsymbol{RV}_{t-5:t-2} + \gamma_m \boldsymbol{W} \cdot \boldsymbol{RV}_{t-22:t-6} + \boldsymbol{u}_{t},\\
    = \boldsymbol{\alpha} &+ \boldsymbol{V}_{:t-1} \boldsymbol{\beta} + \boldsymbol{W} \boldsymbol{V}_{:t-1} \boldsymbol{\gamma} + \boldsymbol{u}_{t},
\end{aligned}
\vspace{-2mm}
\end{equation}%
where $\boldsymbol{\alpha} \in \mathbb{R}^{N}, \boldsymbol{\beta}, \boldsymbol{\gamma} \in \mathbb{R}^{3}$ are parameters to be estimated. $\boldsymbol{W}= \boldsymbol{O}^{-\frac{1}{2}} \boldsymbol{A} \boldsymbol{O}^{-\frac{1}{2}}$ is the normalized adjacency matrix without self-connections, where $\boldsymbol{O}=\operatorname{diag}\left\{n_{1}, \ldots, n_{N}\right\}$ and $n_i = \sum_{j} \boldsymbol{A}[i,j], \, \forall i$.\footnote{It is worth noting that for GHAR, the normalization of $\boldsymbol{W}$ does not impact the forecasting performance directly. However, it does assist with evaluating the relative effect of 0th-hop neighbors in comparison to 1st-hop neighbors.}

\cite{GHAR_zhang2022graph} constructed different types of graphs and concluded that adjacency matrices obtained through Graphical LASSO effectively capture the relationships between individual volatilities, thereby enhancing forecasting accuracy. 

\paragraph{Graphical LASSO} (GLASSO), proposed by \cite{friedman2008sparse}, is a sparsity-penalized maximum likelihood estimator for the precision matrix ${\boldsymbol{\Theta}}$  (i.e. the inverse of the covariance matrix). It assumes the input vector of $N$ nodes is drawn from a multivariate Gaussian distribution $\mathcal{N}(\mathbf{0}, \boldsymbol{\Sigma})$. If the $ij$-th entry of the precision matrix is zero, the returns of $i$-th asset and $j$-th asset are conditionally independent. Therefore, the adjacency matrix $\boldsymbol{A}$ from GLASSO is defined as $\boldsymbol{A}[i,j] = 1$ if ${\boldsymbol{\Theta}}[i,j] \neq 0$; otherwise $\boldsymbol{A}[i,j] = 0$. 

One key advantage of Graphical LASSO is its ability to estimate the conditional independence of assets based on historical returns. Additionally, it offers stability in estimation even in high-dimensional settings where the number of assets exceeds the number of returns. Based on these compelling results, we adopt Graphical LASSO to establish the volatility graph in our study.

%%%%%%%%%%%%%%%%%%%%%%%%%%%%%%%%%%%%%%%%%%%%%%%%%%%%%%%%%%%%%%
%%%%%%%%%%%%%%%%%%%%%%%% Graphs %%%%%%%%%%%%%%%%%%%%%%%%%%%%%%
%%%%%%%%%%%%%%%%%%%%%%%%%%%%%%%%%%%%%%%%%%%%%%%%%%%%%%%%%%%%%%
\section{Proposed methodology}\label{sec:proposed_models}

To investigate the presence of multi-hop and nonlinear effects in modeling volatility spillover, we propose a new class of forecasting models based on the GNNs in Section \ref{sec:propose_gnnhar}. Furthermore, we highlight the significance of using various criteria for the estimation of model coefficients in Section \ref{sec:estimation_criterion}. In Section \ref{sec:forecast_loss}, we introduce the forecast evaluation methods and emphasize the differences between estimation criteria and forecast evaluations. 

\subsection{GNN-enhanced HAR (GNNHAR)} \label{sec:propose_gnnhar}

As introduced in \eqref{eq:ghar_1hop}, GHAR in \cite{GHAR_zhang2022graph} assumes a linear relationship between the volatilities of two connected assets. However, if the spillover effect is nonlinear, linear models are misspecified and are likely to generate less accurate forecasts. Additionally, GHAR considers only the 0th-hop and 1st-hop neighbors, and this lack of consideration for multi-hop neighbors may lead to incomplete information and less accurate predictions. In light of the abilities of GNNs discussed in Section \ref{sec:gnn_overviews}, we propose the following GNN architecture for modeling the volatility spillover effect, allowing for nonlinearity and multi-hop neighbors to improve prediction accuracy. 

% Initially developed for social, citation, and molecular networks, to give a few examples, \MC{out-of-the-box?} GNN architectures may not be suitable for volatility modeling. 
% While various GNN models have shown success on high-dimensional data such as images and videos, they may overfit when applied to financial data with a low signal-to-noise ratio. 
% Additionally, due to the high edge density in financial graphs, noise in individual stocks can propagate to other stocks, negatively affecting their representations and the overall model accuracy.
% % with an average edge density of 0.51 compared to 0.0005 in citation networks investigated by the GCN and GIN models (\cite{gcn_Kipf2017GCN, gnn_gin_xu2018powerful}). 
% To overcome these issues, we propose a custom GNN propagation rule 
\begin{equation}
\label{eq:GNN_layer}
    \textbf{GNN}(\boldsymbol{H}^{(l)}, \boldsymbol{A}): \quad \boldsymbol{H}^{(l+1)} =  \operatorname{ReLU} \left( \boldsymbol{O}^{-\frac{1}{2}} \boldsymbol{A} \boldsymbol{O}^{-\frac{1}{2}} \boldsymbol{H}^{(l)} \boldsymbol{\Theta}^{(l)} \right),
\end{equation}%
where $\boldsymbol{W} = \boldsymbol{O}^{-\frac{1}{2}} \boldsymbol{A} \boldsymbol{O}^{-\frac{1}{2}}$ is the normalized adjacency matrix, used to avoid numerical instabilities and exploding/vanishing gradients during the training phrase. Note that $\boldsymbol{H^{(0)}} = \boldsymbol{V}_{:t-1} \in \mathbb{R}^{N \times 3}$, which is the matrix composed of the past daily, weekly and monthly volatilities. $\boldsymbol{H}^{(l)} \in \mathbb{R}^{N \times D^{(l)}}$ is a matrix of node representations at the $l$-th layer of GNN, where $D^{(l)}$ is the dimension of node representations. $\boldsymbol{\Theta}^{(l)} \in \mathbb{R}^{D^{(l)} \times D^{(l+1)}}$ is a matrix of trainable parameters. \par 

\begin{figure}[H]
    \centering
    \includegraphics
    [width=1.0\textwidth]{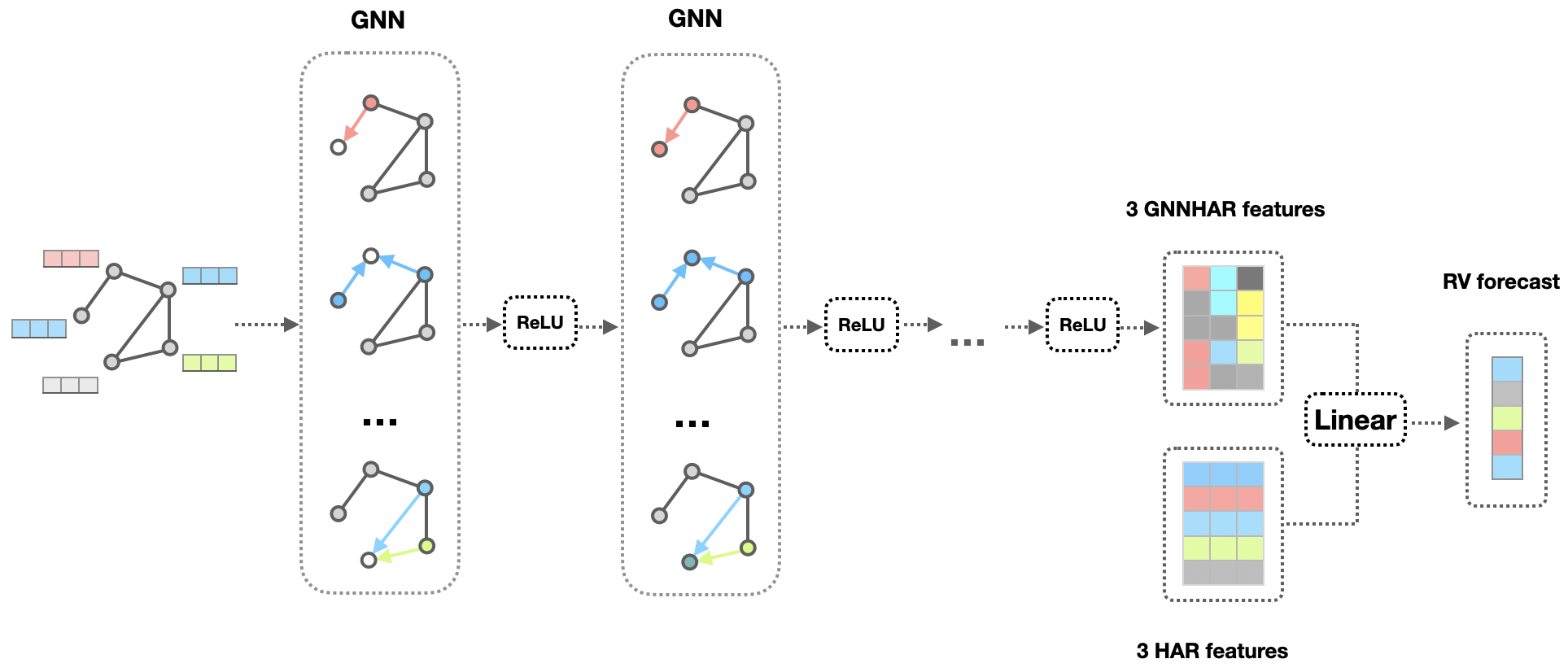}
    \caption{An illustration of the GNNHAR model.}
    \label{fig:illustration_gnnhar}
\end{figure}

In contrast to the GCN architecture shown in \eqref{eq:gnn_literautre_gcn}, our proposed GNN propagation rule does not include self-connections, i.e. the diagonal elements in $\boldsymbol{A}$ are zeros. 
% Additionally, our model is different from GIN, which shares the coefficients for the diagonal and off-diagonal parts ($\boldsymbol{\Theta}^{(l)}$ in \eqref{eq:gnn_literature_gin}). 
We conjecture that the mechanism of an individual stock's past volatility on its future volatility differs from the spillover effect. As a result, we apply the above GNN propagation in \eqref{eq:GNN_layer} solely to model the spillover effect, while the impact of a stock's own past volatility is modeled using the same linear model as in HAR.\footnote{For a study on the non-linearity between a stock's past volatility and its future volatility, we refer readers to \cite{zhang2022volatility}, which suggested that introducing nonlinearity does not result in additional predictive power when modeling daily RV.} This allows for a clear and straightforward explanation of the performance gain of our proposed model compared to the baseline models, HAR and GHAR.

% Specifically, we expect our method to explain whether there is any non-linearity in the volatility spillover effect. To achieve this goal, we need to disentangle the potential influence of nonlinearity between one's own past and future volatility.\footnote{We refer readers to \cite{zhang2022volatility} for a study on the non-linearity between a stock's past volatility and its future volatility.} As a result, the performance gain of our model compared to the baselines HAR and GHAR can be clearly and simply explained.

% By replacing the linear neighborhood aggregation in GHAR, namely the term $\boldsymbol{WV}_{:t-1}\boldsymbol{\gamma}$ in \eqref{eq:ghar_1hop}, with the proposed GNN layer in \eqref{eq:GNN_layer}, we develop a GNN-enhanced HAR model, denoted as GNNHAR1L, as shown in \eqref{eq:gnnhar}. It is worth noting that the main difference between GNNHAR1L and GHAR is the use of the graph convolutional layer with a nonlinear activation function in GNNHAR1L.

We introduce a GNN-enhanced HAR model, referred to as GNNHAR1L in \eqref{eq:gnnhar}, by replacing the linear neighborhood aggregation in GHAR (i.e. the term $\boldsymbol{WV}_{:t-1}\boldsymbol{\gamma}$ in \eqref{eq:ghar_1hop}) with the proposed GNN layer in \eqref{eq:GNN_layer}. It is worth noting that the main difference between GNNHAR1L and GHAR is that GNNHAR1L uses a graph convolutional layer with a nonlinear activation function, in the form of 
\begin{equation}\label{eq:gnnhar}
\begin{aligned}
\boldsymbol{H}^{(1)} & = \textsc{GNN}(\boldsymbol{V}_{:t-1},~ \boldsymbol{A}) \\ 
\textbf{GNNHAR1L}(\boldsymbol{A}):\quad \boldsymbol{RV}_{t} &= \boldsymbol{\alpha} + \boldsymbol{V}_{:t-1} \boldsymbol{\beta} + \boldsymbol{H}^{(1)}\boldsymbol{\gamma} + \boldsymbol{u}_{t}.
\end{aligned}
\end{equation}

As introduced in Section \ref{sec:gnn_overviews}, the nonlinear multi-hop effects can be explored by stacking multiple layers of GNN. We denote the 2-layer and 3-layer models as GNNHAR2L and GNNHAR3L respectively.\footnote{Furthermore, we introduce a linear model that incorporates multi-hop neighbors for volatility forecasting. Additional results regarding this model can be found in Appendix \ref{sec:ghar2hop}.} Specifically, 
\begin{equation} \label{eq:gnnhar2L}
\begin{aligned}
\boldsymbol{H}^{(2)} & = \textsc{GNN}(\boldsymbol{H}^{(1)},~ \boldsymbol{A}) \\
\textbf{GNNHAR2L}(\boldsymbol{A}): \quad  \boldsymbol{RV}_{t} &= \boldsymbol{\alpha} + \boldsymbol{V}_{:t-1} \boldsymbol{\beta} + \boldsymbol{H}^{(2)}  \boldsymbol{\gamma} + \boldsymbol{u}_{t}.
\end{aligned}
\end{equation}%

\begin{equation} \label{eq:gnnhar3L}
\begin{aligned}
\boldsymbol{H}^{(3)} & = \textsc{GNN}(\boldsymbol{H}^{(2)},~ \boldsymbol{A}) \\
\textbf{GNNHAR3L}(\boldsymbol{A}): \quad \boldsymbol{RV}_{t} & = \boldsymbol{\alpha} + \boldsymbol{V}_{:t-1} \boldsymbol{\beta} + \boldsymbol{H}^{(3)}  \boldsymbol{\gamma} + \boldsymbol{u}_{t}.
\end{aligned}
\end{equation}%
 
% We empirically find that every node is at most three steps away from any other node in the volatility spillover graphs uncovered by GLASSO. With a 3-layer GNN, the representation of every asset's volatility is guaranteed to incorporate the information from all other assets. There is thus no need to explore beyond the 3-layer GNN.  

Our empirical analysis (deferred to Appendix \ref{sec:data_stats}) indicates that each node in the volatility spillover graphs  for the components of the DJIA30 index, chosen by GLASSO, is connected to other nodes within a maximum of three steps (i.e. the graph has a diameter of length $3$, which is the size of the longest shortest pairwise path distance in the graph).\footnote{Note that the hyperparameter that determines the sparsity of GLASSO graph estimates is chosen by cross-validation on the GLASSO objective function.}
% \MC{so this is the graph diameter, defined as the maximum length of any shortest path between a pair of nodes. For high diameter graphs, it could be justified to explore beyond layer 3.}
Consequently, by employing a 3-layer GNN, we can guarantee that the volatility representation of each asset encompasses information from all other assets. Hence, there is no requirement to investigate beyond a 3-layer GNN. Nevertheless, it is worth noting that for different universes or graphs, the number of GNN layers may need to be re-evaluated according to the distribution of SPDs.

\subsection{Estimation criterion} \label{sec:estimation_criterion}
The standard HAR model described in \eqref{eq:har} is often estimated via ordinary least squares (OLS). In other words, the estimation criterion (EC) for its in-sample training is the MSE. When the errors $u_{i,t}$ in \eqref{eq:har} are independent, homoscedastic, and normally (Gaussian) distributed, the OLS estimator is consistent under the asymptotic sense. Nonetheless, given the stylized facts of RV (such as spikes, heteroskedasticity, and so on), the OLS estimator may not be an ideal choice and a better estimator may be available. For example, \cite{hansen2022should} proved that the likelihood-based estimator is asymptotically efficient, although the likelihood-based estimator can also be vastly inferior if the underlying statistical model is misspecified. \cite{clements2021practical} empirically compared various estimation criteria on HAR and found that simple weighted least squares can yield substantial improvements to the predictive ability of the standard HAR.

Meanwhile, QLIKE has served as a commonly employed metric for estimating traditional econometric models including GARCH. When $\varepsilon_t$ in \eqref{eq:vol_decomp} has a density (typically unknown), we can utilize the conditional likelihood based on normal density to estimate the models. Specifically, assuming $\varepsilon_t \sim \mathcal{N}(0, 1)$, the conditional Gaussian likelihood function after ignoring constants is $ - \frac{1}{T} \sum_{t} \left[\log \left({\sigma}_{t}^2 \right) + {X}_{t}^2 /{\sigma}_{t}^2 \right]$. It was demonstrated by \cite{hall2003inference, fan2014quasi} that the conditional Gaussian QLIKE estimator is always consistent, even when $\varepsilon_t$ deviates from a normal distribution. 

% \footnote{As demonstrated in \cite{patton2011volatility}, MSE and QLIKE are robust to the presence of noise in the volatility proxy.} 

Utilizing the flexibility of neural networks and the stochastic gradient descent algorithms, we are able to investigate whether different estimation criteria would result in disparate model predictions. Specifically, our primary focus revolves around the following estimation criteria: MSE and QLIKE, defined as follows 

\begin{itemize}
    \item {MSE}: 
    \begin{equation}
        \frac{1}{N} \sum_{i=1}^{N} \frac{1}{ \#\mathcal{T}_{train}} \sum_{t\in \mathcal{T}_{train}}\left({RV}_{i, t}-\widehat{{RV}}_{i, t}^{(F)}\right)^{2},
    \end{equation}
    \item {QLIKE}:
    \begin{equation}
        \frac{1}{N} \sum_{i=1}^{N} \frac{1}{\#\mathcal{T}_{train}} \sum_{t\in \mathcal{T}_{train}} \left[ \frac{{RV}_{i, t}}{\widehat{{RV}}_{i, t}^{(F)}} - \log\left(\frac{{RV}_{i, t}}{{\widehat{{RV}}_{i, t}^{(F)}}}\right)-1 \right],
    \end{equation}
\end{itemize}
where $\widehat{{RV}}_{i, t}^{(F)}$ represents the predicted value of ${RV}_{i, t}$ by a specific model $F$. $N$ is the number of stocks in our universe, $\mathcal{T}_{train}$ is the training period, and $\#\mathcal{T}_{train}$ is the length of the training period. 

Lower values are preferred for both measures. For clarity, we will use $F_M$ ($F_Q$) to denote the model $F$ trained with MSE (QLIKE). To the best of our knowledge, adopting QLIKE as the estimation criterion to optimize volatility models, especially those grounded on neural networks, has not yet drawn considerable attention within the literature.

\begin{figure}[H]
    \centering
    \caption{A comparison of the MSE and QLIKE loss functions.} 
    \includegraphics[width=0.54\textwidth]{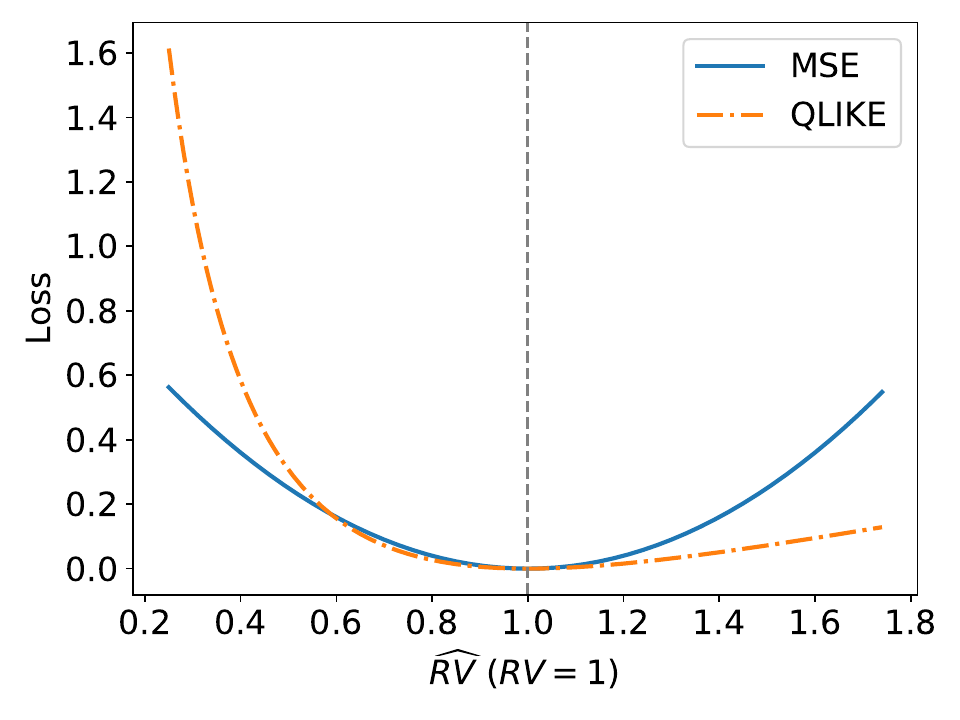}
    % \caption*{\textit{Note:} xxx }
    \label{fig:mse_qlike_loss}
\end{figure}

Figure \ref{fig:mse_qlike_loss} displays the aforementioned EC for different forecasts $\widehat{{RV}}$ when $RV=1$. Notably, the QLIKE function exhibits asymmetry and imposes a higher penalty on under-predictions. This feature becomes particularly significant during turbulent periods, as the volatility forecasts tend to be smaller than the actual shocks. By placing emphasis on under-predictions, models trained with QLIKE have the potential to achieve improved prediction accuracy during such turbulent periods.
% Asymmetric loss in finance
% \XP{Is it the improvement also comes from, in calm period, the volatility is overpredicted in MSE?}

% \begin{theorem}
% Uniqueness of QLIKE.
% \end{theorem}

% \begin{example}
% Restricted case ($\alpha=0$)
% \end{example}

\subsection{Forecast evaluation approaches}\label{sec:forecast_loss}
Regarding the performance of forecasts in out-of-sample tests, we continue to employ MSE and QLIKE as our evaluation methods. However, it is important to distinguish between the concept of forecast loss (FL) and the estimation criterion (EC), as they serve distinct purposes. FL assesses the performance of RV forecasts during out-of-sample testing, while EC is utilized for model estimation within the in-sample period. 

% In terms of out-of-sample evaluation methods, we still adopt the MSE and QLIKE. The concept of forecast loss (FL) is similar yet distinct from EC, as \textit{FL evaluates the performance of RV forecasts in out-of-sample testing while EC is used for in-sample training}. These FL values are computed over the testing period rather than the training period as in EC.

% To verify whether the performance gain beyond the baseline models is significant, we use the following two statistical tests that are widely used in the literature. According to \cite{patton2009evaluating}, QLIKE is more powerful in the Diebold-Mariano (DM) test than MSE. Therefore, we focus more on QLIKE in the out-of-sample results.

In order to determine the significance of the performance improvement compared to the baseline models, we employ two commonly used statistical tests found in the literature. As suggested by \cite{patton2009evaluating}, QLIKE demonstrates greater statistical power than MSE in the Diebold-Mariano (DM) test. Consequently, our focus in the analysis of the out-of-sample results is primarily on QLIKE.
\begin{itemize}
    \item \textbf{Model Confidence Set (MCS)} was proposed by \cite{hansen2011model} to identify a subset of models with significantly superior performance from model candidates, at a given level of confidence. The MCS procedure renders it possible to make statements about the statistical significance from multiple pairwise comparisons. For additional details, we refer to the studies of \cite{hansen2003choosing, hansen2011model}.
    % Consistent with \cite{benchmark2018Symitsi, GHAR_zhang2022graph}, we implement the MCS procedure using a block bootstrap procedure with 10,000 replications and a block length of 2 observations. 
    
    \item \textbf{Diebold-Mariano (DM) test} was proposed by \cite{diebold1995comparing} to examine whether there are significant differences between two time-series forecasts. The DM test was further modified by \cite{harvey1997testing}, to account for serial dependence in forecasts. In addition to comparing errors for each individual stock, we also follow \cite{gu2020empirical} to compare the cross-sectional average of prediction errors from two models. Further details of the DM test are available in  \cite{diebold1995comparing}. 
\end{itemize}%

% {In the main analysis (Section \ref{sec:main_result}), we compare the model performance trained under different evaluation criteria. In the subsequent sections, we focus on the models with MSE as EC.}

% \subsection{{Utility benefits}}

%%%%%%%%%%%%%%%%%%%%%%%%%%%%%%%%%%%%%%%%%%%%%%%%%%%%%%%%%%%%%%%%%%%%%%%%%%%%%%%%%%%%%%%%%%%%%%%%%%%%%%%%%%%%%%%%%%%%%%%%%%%%%%%%%%%%%%%%%%%%%%%%%%%%%%%%%%%%%%%%
\section{Empirical analysis}\label{sec:experiment}
In this section, we first introduce the data and provide details regarding the implementation. Subsequently, we present the main findings and conduct a stratified analysis to evaluate the performance across different market regimes.

% introduce the forecast evaluation techniques and distinguish them from the estimation criterion. Section \ref{sec:main_result} 

\subsection{Setup} \label{sec:experiment_setup}
The intraday data of Dow Jones Industrial Average (DJIA) components are obtained from the LOBSTER database.\footnote{https://lobsterdata.com/} The time period under consideration is from July 1, 2007 to Jun 30, 2021.\footnote{The LOBSTER database contains data from June 27, 2007, up until the day before yesterday} 
Following \cite{bollerslev2016exploiting}, we include only those stocks among the DJIA components that traded continuously throughout the entire period. As a result, 27 stocks are included in the final sample, and their ticker symbols are summarized in Appendix \ref{sec:data_stats}, where we also present summary statistics for the volatility estimates. Additionally, for robustness checks, we consider a larger universe of S\&P100 components. Further details regarding this analysis can be found in Section \ref{sec:robust_large_univ}.

% We also consider a larger universe of S\&P100 components in robustness tests; for further details, see Section \ref{sec:robust_large_univ}. 

Our out-of-sample forecast comparisons are based on the RV forecasts for the set of models introduced in Sections \ref{sec:gnn_overviews} and \ref{sec:proposed_models}. All models are re-calibrated every month based on a rolling sample window of the past 1000 days, following \cite{bollerslev2016exploiting, bollerslev2018modeling, benchmark2018Symitsi, pascalau2021increasing}. Specifically, we use 36-month data for model training, and the recent 12-month data as the validation set to tune the hyperparameters and prevent overfitting.\footnote{To examine the impact of validation dataset, we perform a robustness check for GNNHAR models in Section \ref{sec:robust_valid}, and we conclude that the other choice of validation data does not significantly alter our findings.} Finally, testing data are the samples in the following month; they are out-of-sample in order to provide objective assessments of the model performance. To this end, in aggregate, we obtain a 10-year out-of-sample period, that is, from July 1, 2011 to June 30, 2021.

The parameters in $\text{HAR}_M$ and $\text{GHAR}_M$ are estimated by OLS using both the training and validation data, as there is no requirement for hyperparameter tuning. To estimate the parameters in the proposed GNNHARs, we adopt the Adam optimizer (\cite{gnn_train_adam_kingma2014adam}).\footnote{Adam is a popular stochastic optimization algorithm for deep learning models and is very efficient to find the local minimum, especially with those non-convex and less smooth loss functions.} When QLIKE is chosen as the EC, there are no available estimators in closed form. Therefore, we also employ Adam to optimize $\text{HAR}_Q$ and $\text{GHAR}_Q$ using both the training and validation data. Given the stochastic nature of the optimizer\footnote{The stochastic optimization algorithms might be ended up with different local minima with different initial values.}, we employ an ensemble approach to enhance the robustness of GNNHAR models and QLIKE-trained linear models (see \cite{gu2020empirical, zhang2022volatility}). We train multiple models with random initialization and obtain final predictions by averaging the outputs of all networks. For further details on the hyperparameter choices in GNNHAR, please refer to Appendix \ref{sec:hyperparameter_GNN}.

One-day forecasting is not the only time horizon of interest to practitioners. Following the convention established in the literature (\cite{benchmark2018Symitsi, GHAR_zhang2022graph}), we also examine whether the proposed methods can be applied to various forecasting horizons, e.g. one week or one month. The weekly and monthly target volatility are defined as $\boldsymbol{RV}_{t:t+h} = \sum_{k=0}^{h}\boldsymbol{RV}_{t+k}$, where $h = 4$ and 21, respectively. 

\subsection{Main results}\label{sec:main_result}
We begin our empirical analysis by comparing the out-of-sample performance of the competing models under consideration. Table \ref{tab:rv_main} presents the ratio of forecast losses for each model relative to the $\text{HAR}_M$ model (i.e. HAR estimated by OLS).

Table \ref{tab:rv_main} first highlights the consistent improvement of the GHAR model over the standard HAR model in both forecast losses (FL), implying the importance of graph information. Furthermore, the first two columns of Table \ref{tab:rv_main}, which represent results for the 1-day horizon, demonstrate that our proposed GNNHAR model with a single hidden layer ($\text{GNNHAR1L}_M$), further improves the performance of the linear model $\text{GHAR}_M$. This finding underscores the significance of incorporating nonlinearity when modeling the spillover effect. However, it is worth noting that the performance starts to decline when additional GNN layers are added, particularly with three layers.

\begin{table}[H]
    \centering
    \caption{Out-of-sample forecast losses.}
    \resizebox{0.75\textwidth}{!}{\begin{tabular}{lllllll}
\toprule
& \multicolumn{2}{c}{{1-Day}} & \multicolumn{2}{c}{{1-Week}} & \multicolumn{2}{c}{{1-Month}}  \\
     \cmidrule(lr){2-3}\cmidrule(lr){4-5}\cmidrule(lr){6-7}
                & MSE    & QLIKE  & MSE    & QLIKE  & MSE     & QLIKE  \\\midrule
HAR$_M$          & 1.000  & 1.000  & 1.000  & 1.000  & 1.000   & 1.000  \\
GHAR$_M$         & 0.927  & 0.983  & 0.904  & 0.987  & 0.975$^*$   & 1.036  \\
GNNHAR1L$_M$     & 0.907  & 0.979  & 0.940  & 0.943  & 1.021   & 0.968  \\
GNNHAR2L$_M$     & 0.967  & 0.977  & 1.034  & 0.953  & 1.134   & 1.032  \\
GNNHAR3L$_M$     & 1.210  & 0.982  & 1.014  & 0.961  & 1.046   & 0.958  \\
HAR$_Q$          & 0.927  & 0.981  & 0.939  & 0.945  & 1.069   & 0.986  \\
GHAR$_Q$         & 0.886  & 0.983  & 0.842$^*$ & 0.936  & 1.151   & 0.954$^{\dagger}$ \\
GNNHAR1L$_Q$     & 0.867$^*$ & 0.961$^{\dagger}$ & 0.855  & 0.913$^*$ & 1.179   & 0.965  \\
GNNHAR2L$_Q$     & 0.879  & 0.959$^*$ & 0.873  & 0.920  & 1.736   & 0.947$^*$ \\
GNNHAR3L$_Q$     & 0.894  & 0.963  & 1.185  & 0.942  & 1.502   & 0.971  \\\bottomrule
\end{tabular}
}
    \label{tab:rv_main}
    \caption*{\textit{Note:} The table reports the ratios of forecast losses of various models compared to the standard $\text{HAR}_{M}$ model over the 1-day, 1-week, and 1-month horizons, respectively. The model with the lowest average out-of-sample loss is marked with an asterisk (*). A dagger ($\dagger$) indicates models that yield as accurate forecasts as the best model at the 5\% significance level based on the Model Confidence Set (MCS) test.}
\end{table}

When considering models trained with QLIKE, the results for the 1-day horizon reveal that $\text{HAR}_{Q}$ achieves better forecasts than its counterpart $\text{HAR}_{M}$. $\text{GNNHAR1L}_{Q}$ further improves the predictive accuracy of $\text{GNNHAR1L}_{M}$ and yields the best (resp. second best) out-of-sample performance in terms of MSE (resp. QLIKE). Specifically, at the daily forecast horizon, $\text{GNNHAR1L}_{Q}$ has about 13\% (resp. 4\%) lower average forecast error in MSE (resp. QLIKE) compared to the standard $\text{HAR}_{M}$ model. In addition, the MCS test indicates that both $\text{GNNHAR1L}_{Q}$ and $\text{GNNHAR2L}_{Q}$ are included in the subset of best models, based on the QLIKE forecast loss. Interestingly, $\text{GNNHAR3L}_{Q}$ delivers worse out-of-sample performance than GNNs with one or two layers, yet still outperforms its counterpart trained with MSE. These findings suggest that QLIKE might serve as a more effective in-sample estimation criterion than MSE. In the subsequent sections, we will provide further analysis to delve into these results.

The results for weekly and monthly horizons presented in Table \ref{tab:rv_main} demonstrate that models incorporating graph information (including GHAR and various GNNHAR models) exhibit significantly superior forecast accuracy compared to the HAR model over longer horizons, up to one week. Specifically, when examining the QLIKE loss for the 1-week forecast horizon, we observe that $\text{GNNHAR1L}_{Q}$ achieves the best out-of-sample performance. However, as the prediction horizon extends, the ratios approach or even exceed one, particularly for MSE. This suggests that longer-term forecasting becomes less sensitive to graph information. Additionally, we notice that the discrepancy between the ratios based on MSE and QLIKE becomes more pronounced over longer horizons. One possible explanation is that the QLIKE loss is generally less impacted by extreme observations in the testing samples (see \cite{patton2011volatility}). This is particularly relevant considering that such extreme observations may occur more frequently over longer horizons.

\subsection{Market regimes}
% \red{\large Market regimes first! Then analyze the effect of nonlinearity and QLIKE together!!!}

To assess the stability of performance across different market regimes, we perform a stratified out-of-sample analysis on two sub-samples: relatively calm periods when the RV of the S\&P500 ETF index is below the 90\% quantile of its entire sample distribution, and the turbulent periods when the RV is above its 90\% quantile (see \cite{pascalau2021increasing, GHAR_zhang2022graph}). 

The results presented in Table \ref{tab:rv_1day_regime} demonstrate that the enhancements achieved through the introduction of nonlinearity and the selection of QLIKE as the EC are generally consistent across different market regimes. Specifically, when considering calm days and the daily forecast horizon, the models $\text{GNNHAR1L}_M$ and $\text{GNNHAR2L}_M$ appear to be the most effective based on the MSE loss. On the other hand, when evaluating accuracy in terms of QLIKE, the models $\text{GNNHAR1L}_Q$ and $\text{GNNHAR2L}_Q$ provide the most precise forecasts. This outcome is expected since the volatility process tends to be more stable during calm periods. Consequently, if the forecast user has a specific preference for a particular loss function, it would be advisable to optimize the model parameters accordingly. In other words, for stationary time series, the alignment of the training loss (i.e. EC) and the testing loss (i.e. FL) may produce improved forecasts.

Nevertheless, when examining turbulent days and the daily forecast horizon, models trained with QLIKE exhibit greater percentage improvements compared to those trained with MSE across both losses. For instance, the average forecast MSE (QLIKE) loss of $\text{GNNHAR1L}_Q$ is approximately 13\% (2\%) lower than $\text{GNNHAR1L}_M$. This suggests that models trained with QLIKE may possess unique characteristics distinct from their MSE-trained counterparts during turbulent periods. This intriguing discovery will be further explored and analyzed in the subsequent section.

In addition, when considering longer forecast horizons and periods of calmness, $\text{GNNHAR1L}_M$ produces significantly more accurate out-of-sample forecasts relative to other models in terms of MSE. Regarding the QLIKE accuracy, $\text{GNNHAR1L}_Q$ outperforms other models for the weekly horizon, while $\text{GNNHAR2L}_M$ emerges as the top-performing model for the monthly horizon. When transitioning to the volatile periods, we continue to observe the superiority of QLIKE-trained models (especially $\text{GHAR}_Q$) over MSE-trained models, with the exception being the monthly forecast horizon and considering MSE as the FL.

% When turning to the volatile periods, linear models ($\text{GHAR}_M$ and $\text{GHAR}_Q$) outperform nonlinear models. 

% Moreover, by comparing Panel A and B, we observe most improvements are achieved during turbulent periods.

% $\text{GNNHAR1L}_Q$ and $\text{GNNHAR1L}_Q$ appear to be the best model across different market regimes in general. More interestingly, the larger percentage improvements of GNNHAR stem from both MSE and QLIKE \red{Unclear!!!!}. This indicates that GNNHAR exhibits more predictive power over the turbulent period, which is a difficult regime. As previously discussed, the only difference with respect to GHAR is the use of the graph convolutional layer with a nonlinear activation function in GNNHAR1L. Therefore, we conjecture that the  striking performance gain of GNNHAR1L during turbulent periods is due to its ability to better capture and model  the potentially complex interactions among   assets.\footnote{\cite{choudhry2016stock} revealed strong nonlinear spillover effects among the US and Canada or Japan, during the crisis period.} 

\begin{table}[H]
    \centering
    \caption{Stratified out-of-sample forecast losses.}
    \resizebox{0.75\textwidth}{!}{\begin{tabular}{lllllll}
\toprule
 & \multicolumn{2}{c}{{1-Day}} & \multicolumn{2}{c}{{1-Week}} & \multicolumn{2}{c}{{1-Month}}  \\
     \cmidrule(lr){2-3}\cmidrule(lr){4-5}\cmidrule(lr){6-7}
                 & MSE    & QLIKE  & MSE    & QLIKE  & MSE     & QLIKE  \\\midrule
                 & \multicolumn{6}{c}{{Panel A: Bottom 90\%}}  \\ \midrule
HAR$_M$          & 1.000   & 1.000  & 1.000  & 1.000  & 1.000   & 1.000  \\
GHAR$_M$         & 0.961   & 0.998   & 0.949  & 1.001   & 0.967   & 1.027   \\
GNNHAR1L$_M$     & 0.943$^*$  & 0.998   & 0.883$^*$ & 0.960$^{\dagger}$  & 0.923$^*$  & 0.924$^{\dagger}$ \\
GNNHAR2L$_M$     & 0.944$^{\dagger}$ & 0.990   & 0.901  & 0.954$^{\dagger}$ & 0.946$^{\dagger}$ & 0.921$^*$  \\
GNNHAR3L$_M$     & 0.957   & 0.987   & 0.911  & 0.965   & 0.937$^{\dagger}$ & 0.930$^{\dagger}$ \\
HAR$_Q$          & 1.010   & 0.984   & 1.005  & 0.955$^{\dagger}$ & 1.159   & 0.942$^{\dagger}$ \\
GHAR$_Q$         & 0.989   & 1.007   & 1.076  & 1.001   & 1.257   & 1.084   \\
GNNHAR1L$_Q$     & 0.967   & 0.978$^*$  & 0.944  & 0.943$^*$  & 1.478   & 0.977   \\
GNNHAR2L$_Q$     & 0.976   & 0.979$^{\dagger}$ & 0.985  & 0.947$^{\dagger}$ & 1.433   & 0.973   \\
GNNHAR3L$_Q$     & 0.970   & 0.980$^{\dagger}$ & 1.062  & 0.957   & 1.662   & 0.969   \\\midrule

                 & \multicolumn{6}{c}{{Panel B: Top 10\%}}  \\ \midrule
HAR$_M$          & 1.000   & 1.000  & 1.000  & 1.000  & 1.000   & 1.000  \\
GHAR$_M$         & 0.916  & 0.910   & 0.897  & 0.959  & 0.976$^*$  & 1.043  \\
GNNHAR1L$_M$     & 0.895  & 0.903   & 0.949  & 0.908  & 1.033   & 1.007  \\
GNNHAR2L$_M$     & 1.102  & 0.915   & 1.056  & 0.951  & 1.157   & 1.131  \\
GNNHAR3L$_M$     & 1.293  & 0.958   & 1.030  & 0.952  & 1.059   & 0.982  \\
HAR$_Q$          & 0.900  & 0.965   & 0.928  & 0.925  & 1.059   & 1.024  \\
GHAR$_Q$         & 0.852  & 0.867$^{\dagger}$  & 0.804$^*$ & 0.799$^*$ & 1.149   & 0.841$^*$ \\
GNNHAR1L$_Q$     & 0.834$^*$ & 0.879   & 0.841  & 0.848  & 1.143   & 0.955  \\
GNNHAR2L$_Q$     & 0.848  & 0.862$^*$  & 0.924  & 0.861  & 1.773   & 0.886  \\
GNNHAR3L$_Q$     & 0.868  & 0.882   & 1.205  & 0.909  & 1.483   & 0.973  \\\bottomrule
\end{tabular}}
    \label{tab:rv_1day_regime}
    \caption*{\textit{Note:} The table reports stratified losses during trading days with the bottom 90\% (Panel A) and the top 10\% (Panel B) RV of the S\&P500 ETF index over the 1-day, 1-week, and 1-month horizons, respectively. The model with the lowest average out-of-sample loss is marked with an asterisk (*). A dagger ($\dagger$) indicates models that yield as accurate forecasts as the best model at the 5\% significance level based on the MCS test.}
\end{table}

%%%%%%%%%%%%%%%%%%%%%%%%%%%%%%%%%%%%%%%%%%%%%%%%%%%%%%%%%%%%%%%%%%%%%%%%%%%%%%%%%%%%%%%%%%%%%%%%%%%%%%%%%%%%%%%%%%%%%%%%%%%%%%%%%%%%%%%%%%%%%%%%%%%%%%%%%%%%%%%%
\section{Discussion}\label{sec:discuss}
The objective of this section is to examine the reasons behind the superior performance of our proposed GNNHAR models trained with QLIKE. Our analysis begins by investigating the impact of the choice of EC on the predictive accuracy of the models. We then delve into exploring the influence of model nonlinearity, followed by the examination of the predictive information obtained from multi-hop neighbors.

\subsection{Impact of evaluation criterion} \label{sec:impact_EC}
As previously mentioned, QLIKE deals with over- and under-predictions differently, which may account for the overall better performance of QLIKE-trained models compared to MSE-trained models. In light of this observation, we examine the forecast errors ($\widehat{{RV}}_{i, t}^{(F)}-{RV}_{i, t}$) and forecast ratios ($\widehat{{RV}}_{i, t}^{(F)} / {RV}_{i, t}$) over the entire testing period and various sub-periods.\footnote{It is worth noting that the MSE loss is solely dependent on the forecast error, while QLIKE exclusively relies on the forecast ratio, as corroborated by \cite{patton2011volatility}.} 

\begin{figure}[H]
\centering
\caption{Grouped box plots for models trained with MSE or QLIKE.} 
\subfigure[Forecast errors.\label{fig:error}]{
    \includegraphics[width=0.47\textwidth, trim=2mm 2mm 2mm 2mm,clip]{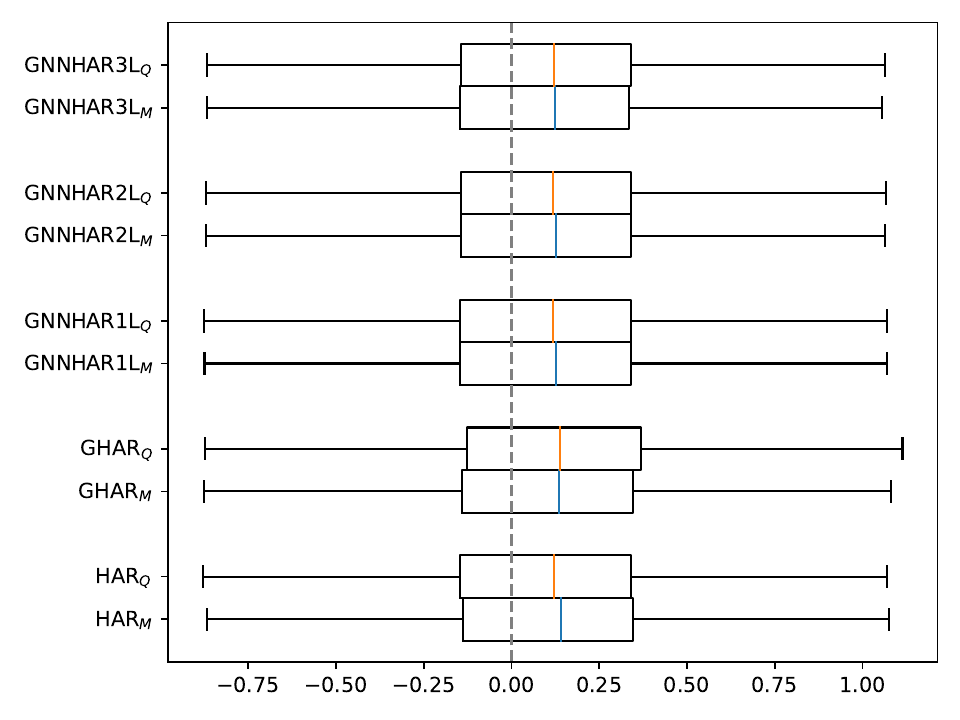}
}
\subfigure[Forecast ratios.\label{fig:ratio}]{
    \includegraphics[width=0.47\textwidth,  trim=2mm 2mm 2mm 2mm,clip]{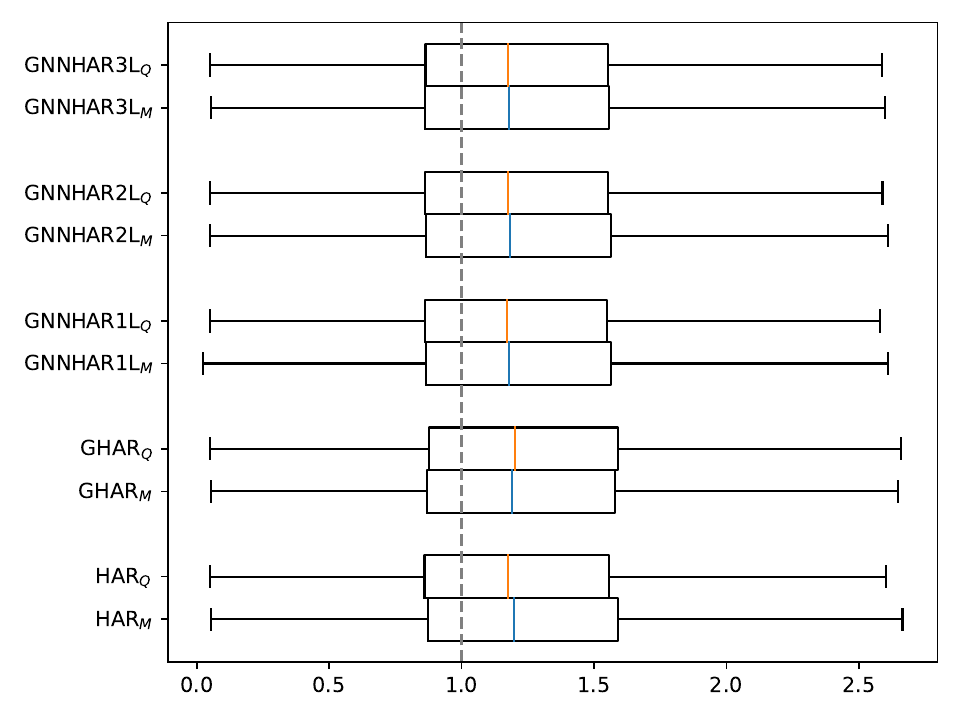}
}
\subfigure[Forecast errors during calm days. \label{fig:low_error}]{
    \includegraphics[width=0.47\textwidth, trim=2mm 2mm 2mm 2mm,clip]{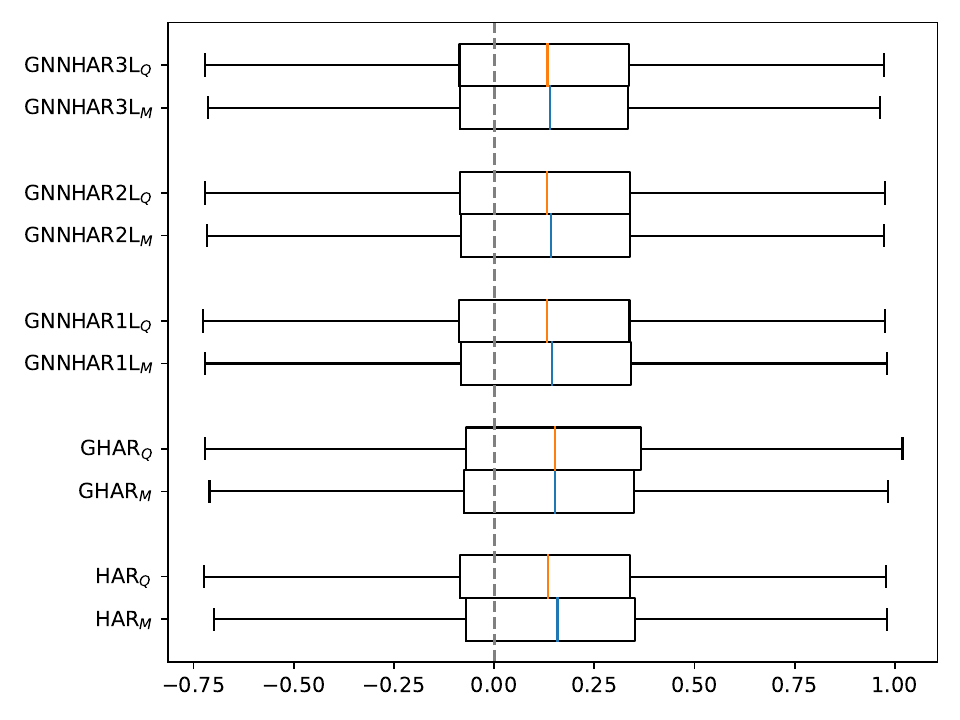}
}
\subfigure[Forecast ratios during calm days.\label{fig:low_ratio}]{
    \includegraphics[width=0.47\textwidth,  trim=2mm 2mm 2mm 2mm,clip]{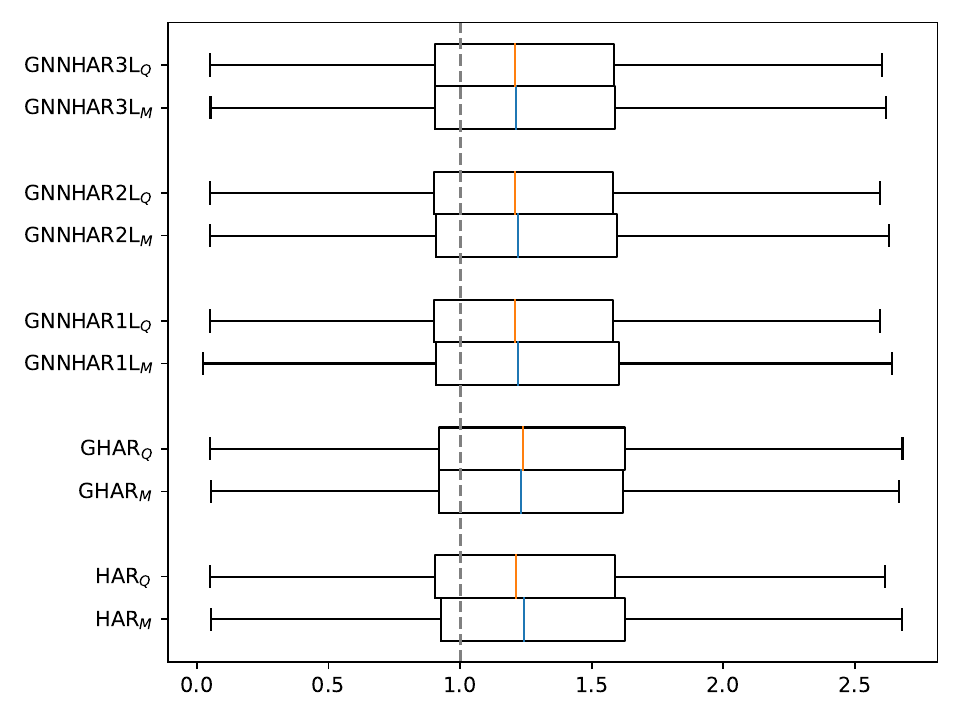}
}
\subfigure[Forecast rrrors during turbulent days.\label{fig:high_error}]{
    \includegraphics[width=0.47\textwidth, trim=2mm 2mm 2mm 2mm,clip]{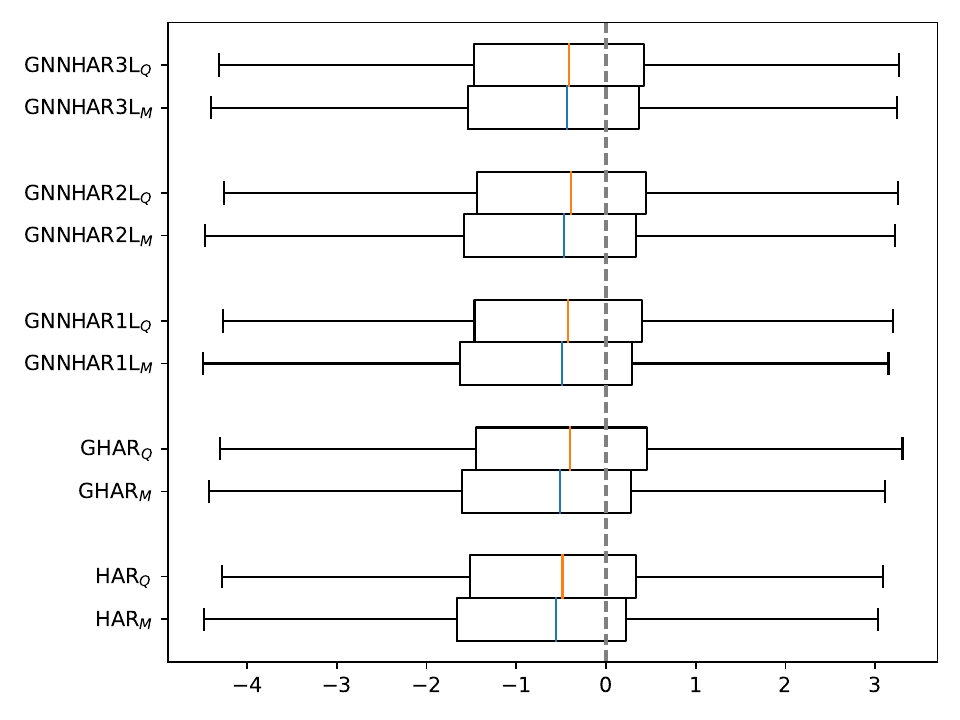}
}
\subfigure[Forecast ratios during turbulent days.\label{fig:high_ratio}]{
    \includegraphics[width=0.47\textwidth,  trim=2mm 2mm 2mm 2mm,clip]{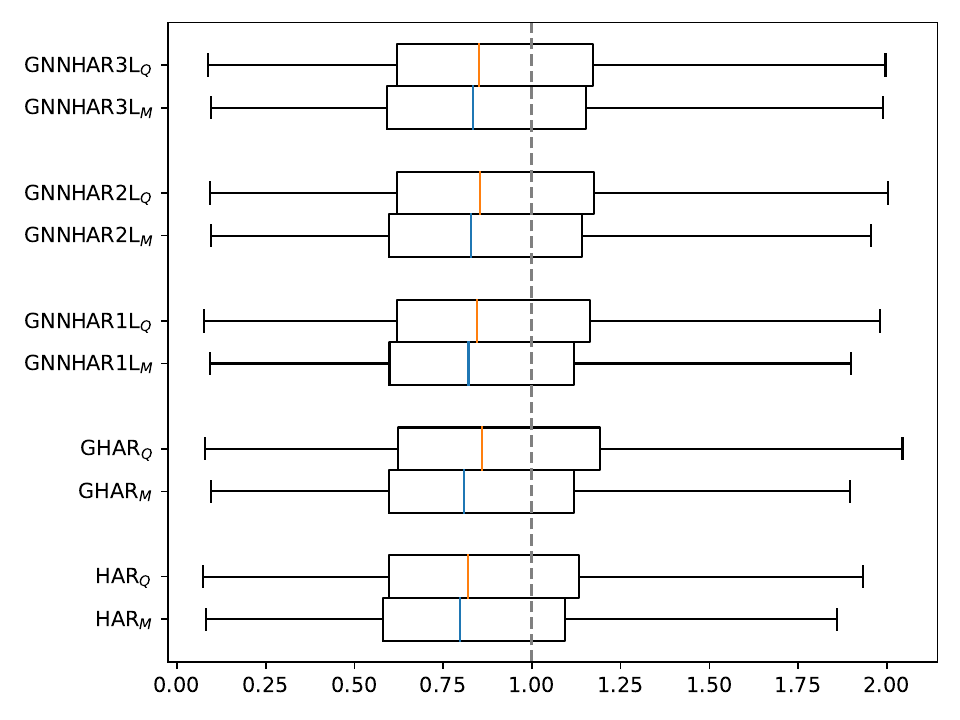}
}
\caption*{\textit{Note:} This figure presents a box plot illustrating five summary statistics: the median, Q1 and Q3 quantiles, and two whiskers. Each group consists of two sets of box plots, with the top (resp. bottom) set representing models utilizing QLIKE (resp. MSE) as EC. (a-b): forecast errors or ratios over the entire testing period. (c-d) forecast errors or ratios over calm periods. (e-f) forecast errors or ratios over turbulent periods.}
\label{fig:error_ratio}
\end{figure}

Figure \ref{fig:error_ratio} presents the box plots for forecast errors and ratios of various models. From subplots (a-b), we observe that in general, all models tend to exhibit a bias towards over-predictions (i.e. positive errors or ratios greater than 1) rather than under-predictions, aligning with the findings of \cite{clements2021practical}. Subplots (c-d) further unveil that this over-prediction tendency is primarily observed during calm periods. Conversely, subplots (e-f) indicate that these models are more inclined to under-predict volatilities during turbulent periods. This observation is not surprising, as the models do not explicitly incorporate any exogenous variables to aid in detecting changes in market conditions. 

Furthermore, subplots (a-b) demonstrate that the bulk of the forecast errors (resp. ratios) of QLIKE-trained models is generally closer to zero (resp. one) compared to MSE-trained models. Specifically, subplots (c-d) reveal that QLIKE-trained models exhibit a reduced tendency to over-predict during calm periods, while subplots (e-f) suggest that they are less prone to excessive under-prediction during turbulent periods, when compared to the MSE-trained models.

\begin{figure}[H]
\centering
\caption{Trajectories of $\beta_d$ in HAR trained with different losses.} 
    \includegraphics[width=0.6\textwidth, trim=1mm 1mm 5mm 1mm,clip]{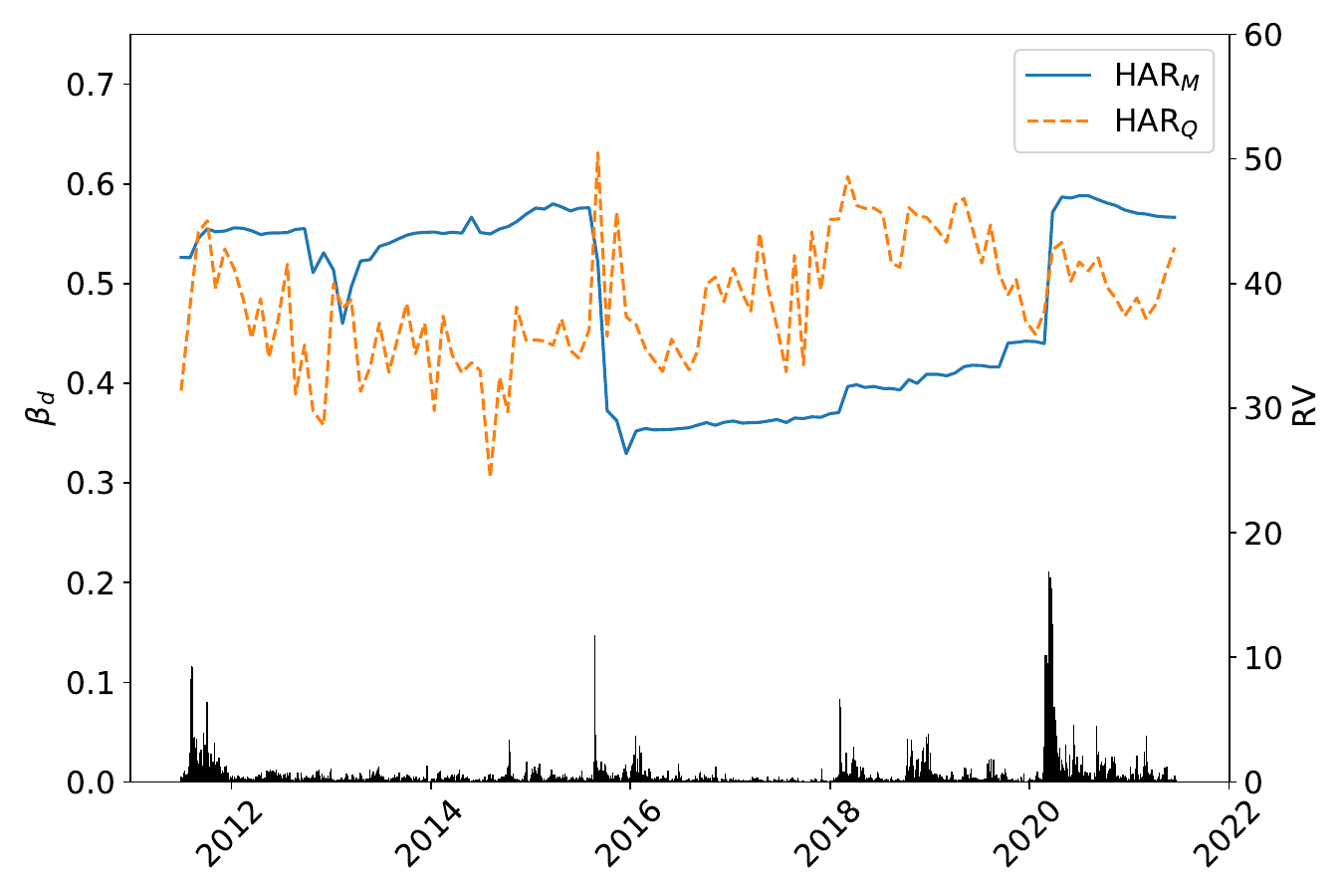}
% \subfigure[HAR$_M$\label{fig:coef_har_m}]{}
% \subfigure[HAR$_Q$\label{fig:coef_har_q}]{
%     \includegraphics[width=0.48\textwidth,  trim=1cm 1mm 5mm 1mm,clip]{figures/HAR_Coef_Itcp_Mean2.pdf}
%  }
\caption*{\textit{Note:} The left $y$-axis represents the estimated values of $\beta_d$ every month, while the right $y$-axis represents the daily RV of S\&P500 ETF shown in bar-charts.}
\label{fig:coef_qlike}
\end{figure}

In order to gain further insights from these findings, we present the trajectories of $\beta_d$ in the HAR models estimated using MSE or QLIKE in Figure \ref{fig:coef_qlike}. As anticipated, there are substantial temporal variations in the rolling estimates of both models. In general, the estimates of $\beta_d$ in $\text{HAR}_Q$ exhibit greater variability compared to those in $\text{HAR}_M$, which can be attributed to the stochastic nature of the optimization algorithm employed in $\text{HAR}_Q$. However, the estimates of $\beta_d$ in $\text{HAR}_M$ reveal two prominent changes occurring during Dec 2015 - Feb 2016 and March 2020 - April 2020, albeit in different directions.\footnote{These two periods correspond to significant market changes, namely the Chinese stock market turbulence and the Covid-19 pandemic.} On the other hand, the patterns of $\beta_d$ in $\text{HAR}_Q$ are comparatively more stable, exhibiting an increasing trend during turbulent periods. This suggests that QLIKE-trained models have the ability to swiftly adapt to market changes and assign greater importance to observations associated with recent significant events. Future studies exploring the relationship between different estimators of HAR are therefore recommended.

\subsection{Impact of nonlinearity}\label{sec:nonlinearity}
% In this section, we first examine whether the model performance varies across different market regimes. This is expected to further provide intuition as to why and when GNNHAR performs as strongly as it appears in Table \ref{tab:rv_main}. Then, we analyze why the nonlinear graph convolution is necessary, with a particular eye on the model interpretability. 

To examine the necessity of nonlinear relations, we provide the following analysis that sheds light on the competitive performance of these models, particularly during volatile periods. Inspired by \cite{chinco2019sparse}, we introduce, for each day $t$, the following metric to evaluate the Fraction of Variance of model $F$ which is Unexplained (FVU) by the standard $\text{HAR}_M$ model\footnote{In fact, $\text{FVU}_t = 1 - R^2(\widehat{RV}_{i, t}^{(F)}, \widehat{RV}_{i, t}^{(\text{HAR}_M)})$, where $R^2$ is the coefficient of determination between the predicted RVs from the target model and the baseline model.} 
\begin{equation}
\text{FVU}_t = \frac{\sum_{i=1}^N \left(\widehat{RV}_{i, t}^{(F)} -\widehat{RV}_{i, t}^{(\text{HAR}_M)} \right)^2} { \sum_{i=1}^N \left(\widehat{RV}_{i, t}^{(F)} - \overline{RV}_{t}^{(F)} \right)^2 },
\end{equation}
where $\overline{RV}_{t}^{(F)}$ is the average forecast RV of model $F$ across stocks on day $t$.
At one extreme, $\text{FVU}_t=0$ means that the $\text{HAR}_M$'s RV forecasts explain all of the variation in the predicted RVs provided by $F$; whereas, at the other extreme, $\text{FVU}_t=1$ denotes that $\text{HAR}_M$ explains none of this variation. 

Table \ref{tab:regime_nonlinear} displays the Fraction of Variance Unexplained (FVU) of each model in comparison to $\text{HAR}_M$. It is worth noting that nonlinear models, particularly those with multiple hidden layers, exhibit higher FVU values, as anticipated. In addition, the results for 1-week and 1-month horizons in Table \ref{tab:regime_nonlinear} suggest that the nonlinearity in volatility models seems to strengthen as the forecasting horizons increase. It is important to mention that the distinction between GHAR and GNNHAR1L lies in the presence of an additional hidden layer with a nonlinear activation function in GNNHAR1L. Consequently, the extra FVUs observed in GNNHAR1L can be considered as a measure of the degree of nonlinearity.

\begin{table}[H]
    \centering
    \caption{FVU compared to HAR$_M$.}
    \resizebox{0.75\textwidth}{!}{\begin{tabular}{lllllll}
\toprule
& \multicolumn{2}{c}{{1-Day}} & \multicolumn{2}{c}{{1-Week}} & \multicolumn{2}{c}{{1-Month}}  \\
     \cmidrule(lr){2-3}\cmidrule(lr){4-5}\cmidrule(lr){6-7}
                 & Bottom & Top   & Bottom & Top    & Bottom & Top  \\\midrule
HAR$_M$          & 0.000  & 0.000 & 0.000  & 0.000 & 0.000   & 0.000 \\
GHAR$_M$         & 0.044  & 0.061 & 0.054  & 0.099 & 0.066   & 0.092 \\
GNNHAR1L$_M$     & 0.077  & 0.165 & 0.117  & 0.244 & 0.178   & 0.300 \\
GNNHAR2L$_M$     & 0.080  & 0.205 & 0.114  & 0.304 & 0.207   & 0.441 \\
GNNHAR3L$_M$     & 0.079  & 0.300 & 0.130  & 0.246 & 0.218   & 0.272 \\
HAR$_Q$          & 0.033  & 0.056 & 0.068  & 0.139 & 0.184   & 0.263 \\
GHAR$_Q$         & 0.077  & 0.128 & 0.108  & 0.216 & 0.228   & 0.779 \\
GNNHAR1L$_Q$     & 0.060  & 0.134 & 0.102  & 0.244 & 0.216   & 0.886 \\
GNNHAR2L$_Q$     & 0.060  & 0.184 & 0.118  & 0.379 & 0.283   & 1.391 \\
GNNHAR3L$_Q$     & 0.070  & 0.212 & 0.163  & 0.764 & 0.292   & 1.236 \\\bottomrule
\end{tabular}

% \begin{tabular}{lcc}
% \toprule
%              & Bottom & Top  \\\midrule
% HAR          & 0.000  & 0.000 \\
% GHAR     & 0.044  & 0.061 \\
% GNNHAR1L & 0.073  & 0.118 \\
% GNNHAR2L & 0.094  & 0.208 \\
% GNNHAR3L & 0.117  & 0.326 \\
% HAR          & 0.000  & 0.000 \\
% GHAR     & 0.044  & 0.061 \\
% GNNHAR1L & 0.073  & 0.118 \\
% GNNHAR2L & 0.094  & 0.208 \\
% GNNHAR3L & 0.117  & 0.326 \\
% \bottomrule
% \end{tabular}
}
    \label{tab:regime_nonlinear}
    \caption*{\textit{Note:} The table reports the fraction of variance unexplained of multiple models compared by the baseline HAR, across different market regimes.}
\end{table}

By comparing the first column and second column in Table \ref{tab:regime_nonlinear}, we observe higher FVU scores during turbulent days, regardless of the choice of EC. This suggests nonlinear spillover effects are most likely to exist in turbulent periods, rather than calm periods. In light of the results in Table \ref{tab:rv_1day_regime}, it can be inferred that a suitable level of model nonlinearity, such as that exhibited by GNNHAR1L, leads to improved predictive power during turbulent days. However, we find that overly complex models, such as GNNHAR3L, are unable to outperform the linear baseline. As a result, GNNHAR1L shows significant promise as a model for capturing nonlinearity, while avoiding the overfitting problem.

\subsection{Impact of multi-hop neighbors} \label{sec:results_multi_hop}

% \subsubsection{Differences between 1-hop and 2-hop models} \label{sec:2Hop-1Hop}

We utilize the DM test to evaluate the statistical significance of 2nd-hop neighbors by comparing the performance of GNNHAR2L and GNNHAR1L. Here, a positive (resp. negative) DM test value indicates the superiority of the GNNHAR1L (resp. GNNHAR2L) model. A $p$-value less than a given significance level $a$ rejects the null hypothesis that GNNHAR2L and GNNHAR1L have the same forecasting power at the $1-a$ confidence level.\footnote{We also conduct the same test to compare linear multi-hop graph models, i.e. GHAR and GHAR2Hop (see Appendix \ref{sec:ghar2hop}) and the conclusions are similar.}

\begin{figure}
    \centering
    \caption{DM test between GNNHAR2L and GNNHAR1L.} 
    \subfigure[$\text{GNNHAR2L}_M \text{ vs } \text{GNNHAR1L}_M$]{
    \includegraphics[width=0.72\textwidth, trim=2cm 1mm 3cm 1.5cm,clip]{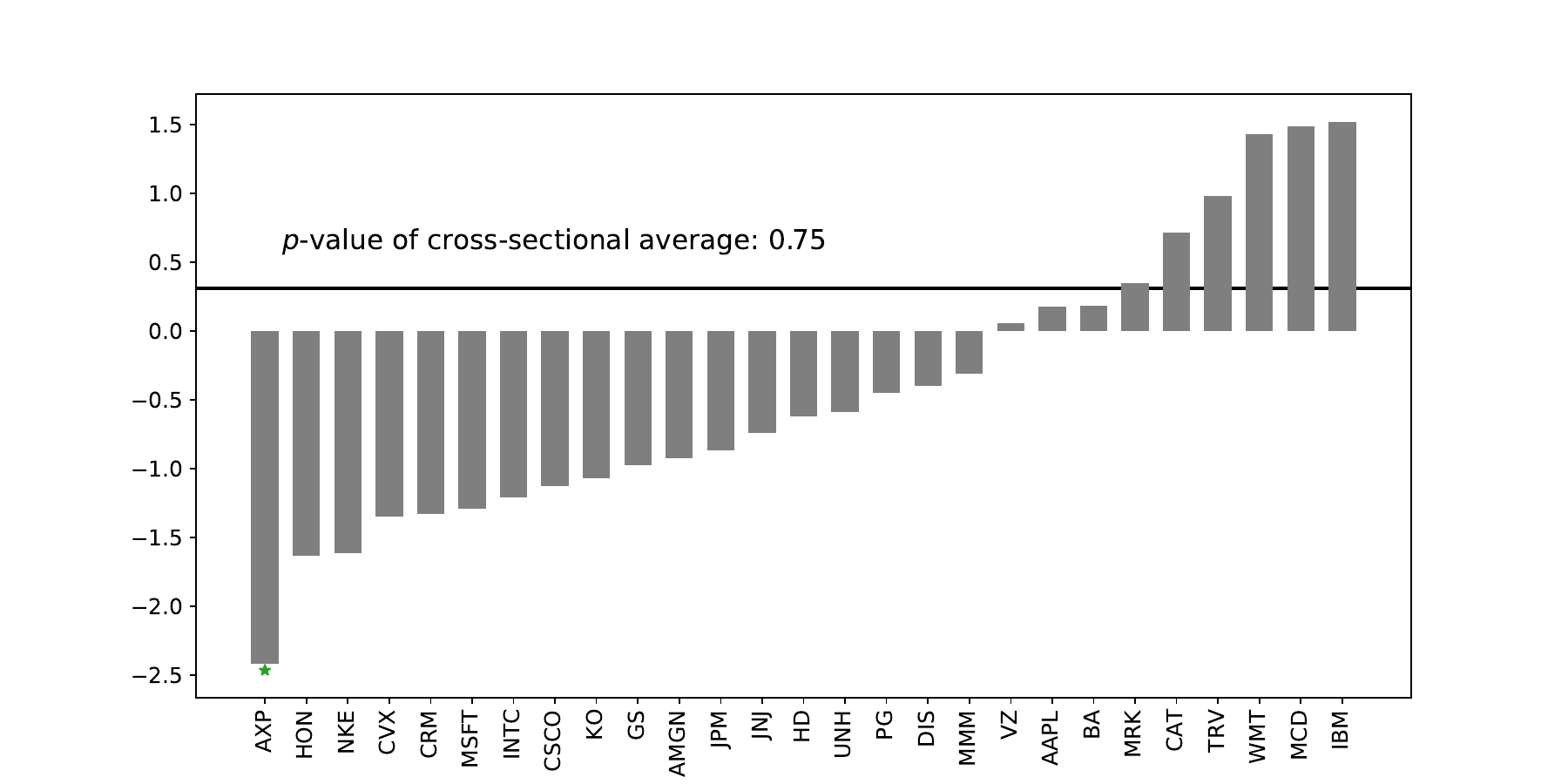}
    }
    \subfigure[$\text{GNNHAR2L}_Q \text{ vs } \text{GNNHAR1L}_Q$]{
    \includegraphics[width=0.72\textwidth, trim=2cm 1mm 3cm 1.5cm,clip]{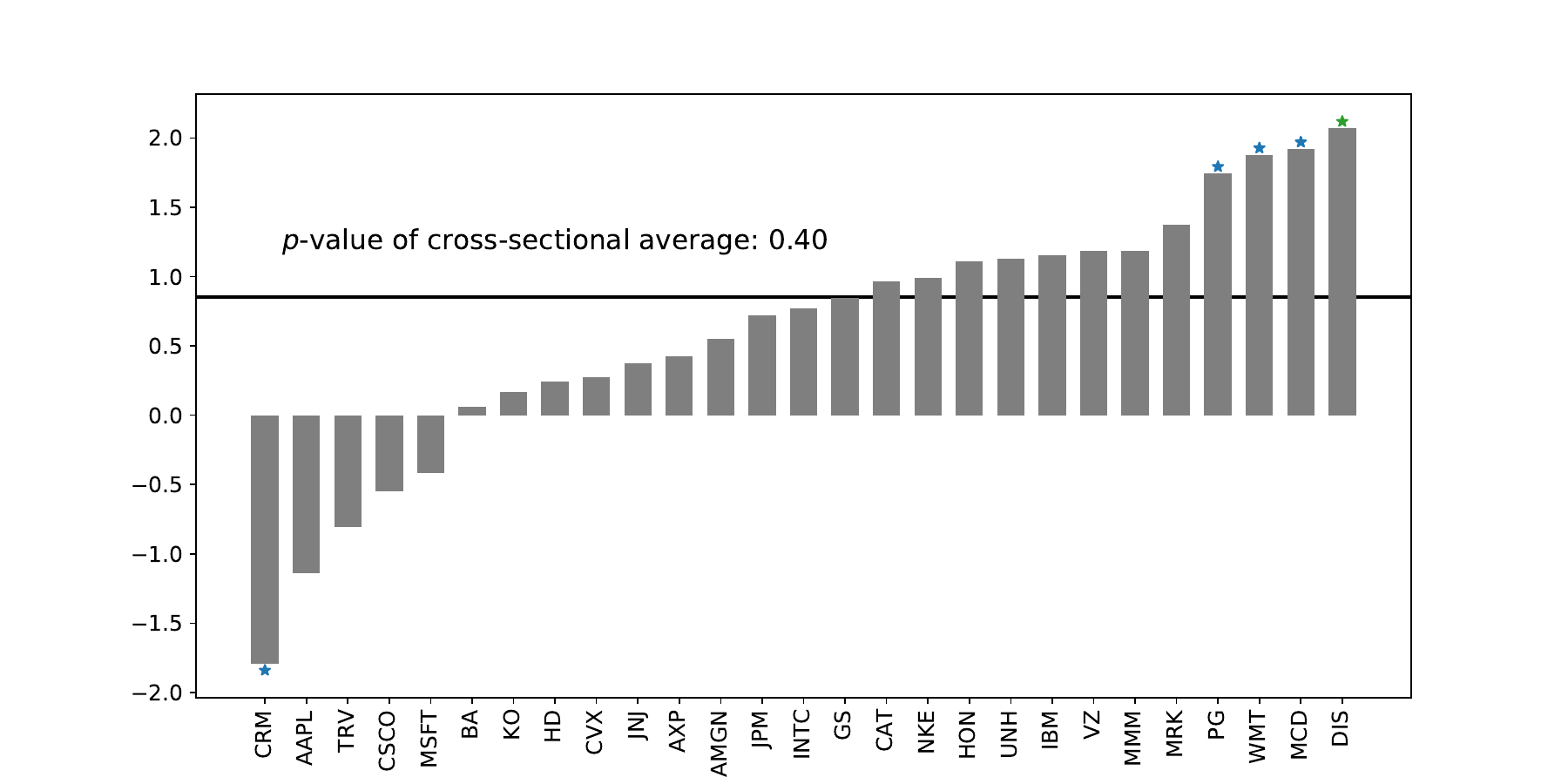}
    }
    \caption*{\textit{Note:} A positive (negative) number indicates superiority for the GNNHAR1L (GNNHAR2L) model. The $y$-axis represents the DM test values based on QLIKE between GNNHAR2L and GNNHAR1L, while the $x$-axis lists the stock symbols. Stars indicate the $p$-value, with orange, green, and blue representing significance at the 1\%, 5\%, and 10\% levels, respectively. The horizon line represents the cross-sectional DM test value and its corresponding $p$-value.}
    \label{fig:DM_GNN}
\end{figure}

Figure \ref{fig:DM_GNN} illustrates the main results from the above hypothesis test. In terms of individual stocks, $\text{GNNHAR2L}_M$ is only superior to $\text{GNNHAR1L}_M$ in forecasting AXP's volatilities, at the 5\% confidence level. When considering the cross-sectional performance, the $p$-value is around 75\%, from which we cannot reject the null hypothesis. This suggests that once the impact from itself and 1st-hop neighbors have been taken into account, 2-hop neighbors are not deemed necessary. The comparison between 
$\text{GNNHAR2L}_Q$ and $\text{GNNHAR1L}_Q$ indeed supports these findings.

% \subsubsection{Over-smoothing in GNNs} \label{sec:over_smoothness}

GNNs are known to suffer from the problem of \textbf{over-smoothing}, which is defined as the high similarity of node representations obtained at the output layer of GNNs, see \cite{gnn_smoothness_li2018deeper}. The high similarity is often observed when stacking with multiple GNN layers that are more than necessary. With  $K$ layers, every node receives information from its $K$-hop neighbors.\footnote{This is also known as the receptive field of GNN. More details have been introduced in Section \ref{sec:gnn_overviews}.} When $K$ is large, node representations obtained from GNN information propagation become indistinguishable and weaken the forecasting accuracy. \par 

Following the convention in the GNN literature (e.g. \cite{gnn_smoothness_chen2020measuring}), we use the Mean Average Distance (MAD) to measure the similarity of node representations and identify whether there is any sign of over-smoothing in our GNNHAR models. MAD takes as input the node representations $\boldsymbol{H} \in \mathbb{R}^{N \times D}$ obtained at the final layer of GNN, that is $\boldsymbol{H} = \text{GNN}(\boldsymbol{V_{:t-1}}, \boldsymbol{A})$ in \eqref{eq:gnnhar}, and is defined as follows 
\footnote{$\boldsymbol{H}$ is the (unweighted) average of the hidden representations obtained from GNNHARs in our ensemble set.}% .}
\begin{equation} \label{eq:oversmoothness}
    \text{MAD} = \frac{\sum_{i=1}^N \bar{d}_i}{\sum_{i=1}^N \mathbbm{1}_{\bar{d}_i> 0} }, ~~ \text{where} ~ \bar{d}_i = \frac{\sum_{j=1}^N \bar{\boldsymbol{D}}_{ij}}{\sum_{j=1}^N \mathbbm{1}_{\bar{\boldsymbol{D}}_{ij > 0}}}.
\end{equation}
$\bar{\boldsymbol{D}}$ is the masked cosine distance matrix, i.e. $\bar{\boldsymbol{D}} = {\boldsymbol{D}} \circ {\boldsymbol{A}}$, where $\circ$ denotes the Hadamard product (element-wise multiplication), and $\boldsymbol{D}_{ij}=1-\frac{\boldsymbol{H}[i, :] \cdot \boldsymbol{H}[j, :]}{\|\boldsymbol{H}[i, :]\|\|\boldsymbol{H}[j, :]\|}$.
% $\boldsymbol{D}_{ij}$ is the $ij$-th entry of the cosine distance $\boldsymbol{D}$ between rows of $\boldsymbol{H}$, i.e. $\boldsymbol{D}_{ij}=1-\frac{\boldsymbol{H}[i, :] \cdot \boldsymbol{H}[j, :]}{\|\boldsymbol{H}[i, :]\|\|\boldsymbol{H}[j, :]\|}$. Note that, we are only interested in connected pairs in GNN propagation, so $\boldsymbol{D}$ is masked by the input graph $\boldsymbol{A}$, i.e. $\boldsymbol{D}_{ij}=0 \text{ iff } \boldsymbol{A}_{ij}=0$.
In the above definition, $\bar{d}_i$ is the average distance between the representations of node $i$ and its connected nodes. Overall, MAD represents an average level of how a node representation is similar to the representations of its connected neighbors in a graph. 
\par

\begin{figure}[H]
    \centering
    \caption{Smoothness of GNNHARs.} 
    \includegraphics[width=0.66\textwidth, trim=3mm 5mm 1cm 1cm,clip]{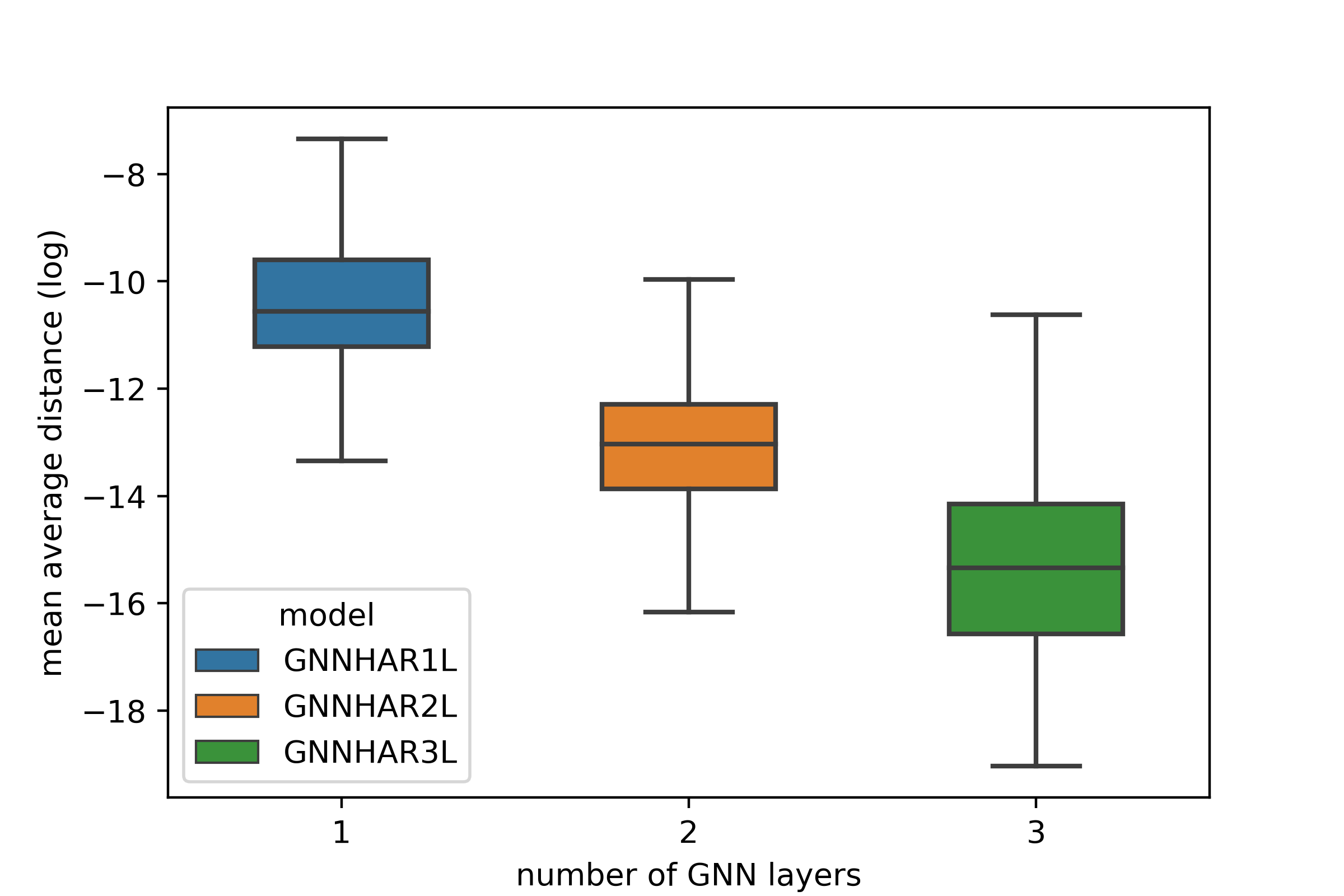}
    \caption*{\textit{Note:} A small value of mean average distance (MAD) indicates a high similarity between node representations at the output layer of GNN. }
    \label{fig:gnn_oversmoothing}
\end{figure}

In Figure \ref{fig:gnn_oversmoothing}, three boxes represent GNNHAR models with 1, 2, and 3 GNN layers trained with MSE.\footnote{Similar results (unreported) are observed for GNNHARs trained with QLIKE.} Each box corresponds to the MAD values on a logarithmic scale, calculated across all out-of-sample samples.  As the number of GNN layers increases, there is a decrease in log MAD that corresponds to an increase in smoothness. The 3-layer GNNHAR has the lowest MAD score, suggesting potential over-smoothing of node representations. Specifically, the rows of $\text{GNN}(\boldsymbol{V_{:t-1}}, \boldsymbol{A})$ from GNNHAR3L in \eqref{eq:gnnhar}, become too similar to provide any node specific predictive information. This partially explains the inferior performance of GNNHAR3L, as shown in Table \ref{tab:rv_main}.

%%%%%%%%%%%%%%%%%%%%%%%%%%%%%%%%%%%%%%%%%%%%%%%%%%%%%%%%%%%%%%%%%%%%%%%%%%%%%%%%%%%%%%%%%%%%%%%%%%%%%%%%%%%%%%%%%%%%%%%%%%%%%%%%%%%%%%%%%%%%%%%%%%%%%%%%%%%%%%%%%%%%%%%%%%%%%%%%%%%%%%%%%%%%%%%%%%%%%%%%%%%%%%%%%%%%%%%%%%%%%%%%%%%%%%%
\section{Robustness tests}\label{sec:robustness}

After presenting the main empirical results and analyzing the model performance across different market periods, we shift our focus to evaluating the robustness of the proposed models by considering two aspects: (i) an alternative validation set size, and (ii) a larger universe.

% To further assess the robustness of our findings and ascertain that they are not specific to the equity markets under considera- tion, we repeat our out-of-sample analys

% In this section, we consider the three aspects regarding the robustness of the proposed models: (i) an alternative validation set size, and (ii) a larger universe.

\subsection{Alternative validation set size}\label{sec:robust_valid}
Our main analysis is based on rolling samples of 4 years, with the first approximately 3 years as training data, and the recent 1 year as validation data. Using a smaller validation data set, such as 1 month, does not significantly alter our findings, as shown in Table \ref{tab:validation_22}.

\begin{table}[H]
    \centering
    \caption{Out-of-sample forecast losses under a smaller validation data set.}
    \resizebox{0.75\textwidth}{!}{\begin{tabular}{lllllll}
\toprule
& \multicolumn{2}{c}{{1-Day}} & \multicolumn{2}{c}{{1-Week}} & \multicolumn{2}{c}{{1-Month}}  \\
     \cmidrule(lr){2-3}\cmidrule(lr){4-5}\cmidrule(lr){6-7}
                & MSE    & QLIKE  & MSE    & QLIKE  & MSE     & QLIKE  \\\midrule
HAR$_M$          & 1.000  & 1.000  & 1.000  & 1.000  & 1.000   & 1.000  \\
GHAR$_M$         & 0.927  & 0.983  & 0.904  & 0.987  & 0.975$^*$   & 1.031  \\
GNNHAR1L$_M$     & 0.942  & 0.978  & 0.931  & 0.945  & 1.008   & 0.975  \\
GNNHAR2L$_M$     & 0.984  & 0.984  & 1.005  & 0.956  & 1.138   & 1.033  \\
GNNHAR3L$_M$     & 1.078  & 1.002  & 1.035  & 0.954  & 1.068   & 0.958  \\
HAR$_Q$          & 0.936  & 0.986  & 0.945  & 0.944  & 1.218   & 0.959  \\
GHAR$_Q$         & 0.942  & 0.982  & 0.993  & 0.945  & 1.174   & 0.954  \\
GNNHAR1L$_Q$     & 0.889$^*$ & 0.967$^*$ & 0.875$^{\dagger}$ & 0.912  & 1.226   & 0.961  \\
GNNHAR2L$_Q$     & 0.896  & 0.968$^{\dagger}$ & 0.861$^*$ & 0.907$^*$ & 1.510   & 0.925$^*$ \\
GNNHAR3L$_Q$     & 1.152  & 0.981  & 1.060  & 0.929  & 1.572   & 0.972   \\\bottomrule
\end{tabular}
}
    \label{tab:validation_22}
    \caption*{\textit{Note:} The table reports the out-of-sample losses of various models using 47 months as training data and the recent 1 month as validation data. The model with the lowest average out-of-sample loss is marked with an asterisk (*). A dagger ($\dagger$) indicates models that yield as accurate forecasts as the best model at the 5\% significance level based on the MCS test.}
\end{table}

\subsection{Larger universe} \label{sec:robust_large_univ}
To further assess the robustness of our findings and ascertain that they are not specific to the stocks under current consideration, we repeat the out-of-sample analysis using a larger data set, including the components of the S\&P100 index.\footnote{Details about the data are provided in Appendix \ref{sec:data_stats}.} The experimental setups and the hyperparameter choices in GNNHAR remain the same as those described in Section \ref{sec:experiment_setup}. As illustrated in Table \ref{tab:spd_freq}, in the volatility spillover graphs for the S\&P100 index components, each node is connected to other nodes within a maximum of 5 steps. Consequently, we extend our analysis to include 4-layer and 5-layer versions of the GNNHAR model.

\begin{table}[H]
    \centering
    \caption{Out-of-sample forecast losses on S\&P100.}
    \resizebox{0.75\textwidth}{!}{\begin{tabular}{lllllll}
\toprule
& \multicolumn{2}{c}{{1-Day}} & \multicolumn{2}{c}{{1-Week}} & \multicolumn{2}{c}{{1-Month}}  \\
     \cmidrule(lr){2-3}\cmidrule(lr){4-5}\cmidrule(lr){6-7}
                & MSE    & QLIKE  & MSE    & QLIKE  & MSE     & QLIKE  \\\midrule
HAR$_Q$          & 1.000  & 1.000  & 1.000  & 1.000  & 1.000   & 1.000  \\
GHAR1L$_Q$       & 0.948  & 0.988  & 0.909  & 0.994  & 0.972$^*$  & 0.986  \\
GNNHAR1L$_Q$     & 0.963  & 0.986  & 0.951  & 0.944  & 1.027   & 1.092  \\
GNNHAR2L$_Q$     & 1.072  & 0.988  & 1.031  & 0.954  & 1.092   & 1.000   \\
GNNHAR3L$_Q$     & 1.061  & 0.986  & 1.029  & 0.959  & 0.992   & 0.967  \\
GNNHAR4L$_Q$     & 1.047  & 0.992  & 1.042  & 0.975  & 1.079   & 0.978  \\
GNNHAR5L$_Q$     & 1.090  & 0.997  & 1.057  & 0.986  & 1.109   & 1.038  \\
HAR$_Q$          & 0.949  & 0.983  & 0.937  & 0.947  & 1.171   & 0.991  \\
GHAR$_Q$         & 0.919  & 0.984  & 0.850$^*$ & 0.922  & 1.154   & 0.939$^*$ \\
GNNHAR1L$_Q$     & 0.917$^{\dagger}$ & 0.969  & 0.858  & 0.916  & 1.231   & 1.017  \\
GNNHAR2L$_Q$     & 0.915$^*$ & 0.969  & 0.909  & 0.915$^*$ & 1.206   & 0.941$^{\dagger}$ \\
GNNHAR3L$_Q$     & 0.938  & 0.966$^*$ & 1.178  & 0.968  & 1.523   & 0.946  \\
GNNHAR4L$_Q$     & 0.985  & 0.970  & 1.165  & 0.972  & 1.563   & 0.971  \\
GNNHAR5L$_Q$     & 0.951  & 0.968  & 1.193  & 0.975  & 1.741   & 0.989  \\
\bottomrule
\end{tabular}
}
    \label{tab:rv_sp100}
    \caption*{\textit{Note:} The table reports the ratios of forecast losses of various models compared to the standard $\text{HAR}_{M}$ model over the 1-day, 1-week, and 1-month horizons, respectively. The model with the lowest average out-of-sample loss is marked with an asterisk (*). A dagger ($\dagger$) indicates models that yield as accurate forecasts as the best model at the 5\% significance level based on the MCS test.}
\end{table}

The out-of-sample forecasting performance on the volatilities of S\&P100 components is presented in Table \ref{tab:rv_sp100}. Firstly, we observe that GHAR consistently enhances forecasting accuracy compared to the traditional HAR model. Additionally, the nonlinear variant, GNNHAR1L, further improves upon the performance of GHAR over the 1-day horizon. Generally, as we increase the number of layers in the GNNHAR models, their forecasting performance tends to decline. Nevertheless, we still observe the benefits of training models with the QLIKE loss function. In summary, the findings presented in Table \ref{tab:rv_sp100} align closely with those observed for DJIA30, providing consistent results across both data sets.

%%%%%%%%%%%%%%%%%%%%%%%%%%%%%%%%%%%%%%%%%%%%%%%%%%%%%%%%%%%%%%%%%%%%%%%%%%%%%%%%%%%%%%%%%%%%%%%%%%%%%%%%%%%%%%%%%%%%%%%%%%%%%%%%%%%%%%%%%%%%%%%%%%%%%%%%%%%%%%%%%%%%%%%%%%%%%%%%%%%%%%%%%%%%%%%%%%%%%%%%%%%%
\section{Conclusion}\label{sec:conclusion}
In this article, we propose a novel methodology GNNHAR for modeling and forecasting RV, while taking into account volatility spillover effects in the U.S. equity market. Our analysis suggests that the information from the multi-hop neighbors in the financial graph does not offer a clear advantage in predicting the volatility of any target stock. However, nonlinear spillover effects help improve the forecasting accuracy of the RV. Moreover, we find that utilizing QLIKE as the training loss function, in comparison to the conventional MSE, leads to more accurate volatility forecasts. Additionally, QLIKE-trained nonlinear models demonstrate greater resilience during turbulent periods compared to calmer market conditions, thereby posing challenges for standard linear models. Our comprehensive evaluation tests and alternative setting confirm the robustness and effectiveness of our proposed methodology.

One interesting direction is to further investigate why utilizing QLIKE instead of MSE, as the evaluation criterion, improves forecasting accuracy. {While \cite{hansen2022should} asserted the asymptotic efficiency of the likelihood-based estimator, our settings differ from theirs in that they assumed the likelihood function is in conjunction with the forecasting loss. Conversely, \cite{patton2011volatility} claimed that MSE is more sensitive to extreme observations than QLIKE, but there is a lack of theoretical underpinnings on how this might improve the predictive powers in various market conditions.}

Another interesting direction to explore is the robustness of the proposed methods when applied to different approaches in constructing financial graphs, such as those based on supply-chain (\cite{herskovic2020firm}) and analyst co-coverage (\cite{ali2020shared}). It would be valuable to investigate whether these graphs provide unique information content and have the potential to enhance performance.\par

\bibliographystyle{plainnat}
\bibliography{ref}

\begin{thebibliography}{60}
\providecommand{\natexlab}[1]{#1}
\providecommand{\url}[1]{\texttt{#1}}
\expandafter\ifx\csname urlstyle\endcsname\relax
  \providecommand{\doi}[1]{doi: #1}\else
  \providecommand{\doi}{doi: \begingroup \urlstyle{rm}\Url}\fi

\bibitem[Acemoglu et~al.(2010)Acemoglu, Ozdaglar, and
  Tahbaz-Salehi]{acemoglu2010cascades}
Daron Acemoglu, Asuman Ozdaglar, and Alireza Tahbaz-Salehi.
\newblock Cascades in networks and aggregate volatility.
\newblock Technical report, National Bureau of Economic Research, 2010.

\bibitem[Ali and Hirshleifer(2020)]{ali2020shared}
Usman Ali and David Hirshleifer.
\newblock Shared analyst coverage: Unifying momentum spillover effects.
\newblock \emph{Journal of Financial Economics}, 136\penalty0 (3):\penalty0
  649--675, 2020.

\bibitem[Alon and Yahav(2020)]{gnn_receptive_field_alon2020bottleneck}
Uri Alon and Eran Yahav.
\newblock On the bottleneck of graph neural networks and its practical
  implications.
\newblock In \emph{International Conference on Learning Representations}, 2020.

\bibitem[Andersen et~al.(2001)Andersen, Bollerslev, Diebold, and
  Ebens]{andersen2001distribution}
Torben~G Andersen, Tim Bollerslev, Francis~X Diebold, and Heiko Ebens.
\newblock The distribution of realized stock return volatility.
\newblock \emph{Journal of Financial Economics}, 61\penalty0 (1):\penalty0
  43--76, 2001.

\bibitem[Andersen et~al.(2011)Andersen, Bollerslev, and
  Meddahi]{andersen2011realized}
Torben~G Andersen, Tim Bollerslev, and Nour Meddahi.
\newblock Realized volatility forecasting and market microstructure noise.
\newblock \emph{Journal of Econometrics}, 160\penalty0 (1):\penalty0 220--234,
  2011.

\bibitem[Bai et~al.(2010)Bai, Wong, and Zhang]{bai2010multivariate}
Zhidong Bai, Wing-Keung Wong, and Bingzhi Zhang.
\newblock Multivariate linear and nonlinear causality tests.
\newblock \emph{Mathematics and Computers in simulation}, 81\penalty0
  (1):\penalty0 5--17, 2010.

\bibitem[Barndorff-Nielsen and Shephard(2002)]{barndorff2002econometric}
Ole~E Barndorff-Nielsen and Neil Shephard.
\newblock Econometric analysis of realized volatility and its use in estimating
  stochastic volatility models.
\newblock \emph{Journal of the Royal Statistical Society: Series B (Statistical
  Methodology)}, 64\penalty0 (2):\penalty0 253--280, 2002.

\bibitem[Bollerslev et~al.(2016)Bollerslev, Patton, and
  Quaedvlieg]{bollerslev2016exploiting}
Tim Bollerslev, Andrew~J Patton, and Rogier Quaedvlieg.
\newblock Exploiting the errors: A simple approach for improved volatility
  forecasting.
\newblock \emph{Journal of Econometrics}, 192\penalty0 (1):\penalty0 1--18,
  2016.

\bibitem[Bollerslev et~al.(2018{\natexlab{a}})Bollerslev, Hood, Huss, and
  Pedersen]{bollerslev2018risk}
Tim Bollerslev, Benjamin Hood, John Huss, and Lasse~Heje Pedersen.
\newblock Risk everywhere: Modeling and managing volatility.
\newblock \emph{Review of Financial Studies}, 31\penalty0 (7):\penalty0
  2729--2773, 2018{\natexlab{a}}.

\bibitem[Bollerslev et~al.(2018{\natexlab{b}})Bollerslev, Patton, and
  Quaedvlieg]{bollerslev2018modeling}
Tim Bollerslev, Andrew~J Patton, and Rogier Quaedvlieg.
\newblock Modeling and forecasting (un) reliable realized covariances for more
  reliable financial decisions.
\newblock \emph{Journal of Econometrics}, 207\penalty0 (1):\penalty0 71--91,
  2018{\natexlab{b}}.

\bibitem[Buncic and Gisler(2016)]{buncic2016global}
Daniel Buncic and Katja~IM Gisler.
\newblock Global equity market volatility spillovers: A broader role for the
  united states.
\newblock \emph{International Journal of Forecasting}, 32\penalty0
  (4):\penalty0 1317--1339, 2016.

\bibitem[Callot et~al.(2017)Callot, Kock, and Medeiros]{callot2017modeling}
Laurent~AF Callot, Anders~B Kock, and Marcelo~C Medeiros.
\newblock Modeling and forecasting large realized covariance matrices and
  portfolio choice.
\newblock \emph{Journal of Applied Econometrics}, 32\penalty0 (1):\penalty0
  140--158, 2017.

\bibitem[Chen et~al.(2020)Chen, Lin, Li, Li, Zhou, and
  Sun]{gnn_smoothness_chen2020measuring}
Deli Chen, Yankai Lin, Wei Li, Peng Li, Jie Zhou, and Xu~Sun.
\newblock Measuring and relieving the over-smoothing problem for graph neural
  networks from the topological view.
\newblock In \emph{AAAI Conference on Artificial Intelligence}, volume~34,
  pages 3438--3445, 2020.

\bibitem[Chen and Robert(2022)]{chen2022multivariate}
Qinkai Chen and Christian-Yann Robert.
\newblock Multivariate realized volatility forecasting with graph neural
  network.
\newblock In \emph{Proceedings of the Third ACM International Conference on AI
  in Finance}, pages 156--164, 2022.

\bibitem[Chen et~al.(2018)Chen, Wei, and
  Huang]{gnn_stock_chen2018incorporating}
Yingmei Chen, Zhongyu Wei, and Xuanjing Huang.
\newblock Incorporating corporation relationship via graph convolutional neural
  networks for stock price prediction.
\newblock In \emph{Proceedings of the 27th ACM International Conference on
  Information and Knowledge Management}, pages 1655--1658, 2018.

\bibitem[Chinco et~al.(2019)Chinco, Clark-Joseph, and Ye]{chinco2019sparse}
Alex Chinco, Adam~D Clark-Joseph, and Mao Ye.
\newblock Sparse signals in the cross-section of returns.
\newblock \emph{Journal of Finance}, 74\penalty0 (1):\penalty0 449--492, 2019.

\bibitem[Choudhry et~al.(2016)Choudhry, Papadimitriou, and
  Shabi]{choudhry2016stock}
Taufiq Choudhry, Fotios~I Papadimitriou, and Sarosh Shabi.
\newblock Stock market volatility and business cycle: Evidence from linear and
  nonlinear causality tests.
\newblock \emph{Journal of Banking \& Finance}, 66:\penalty0 89--101, 2016.

\bibitem[Cipollini et~al.(2020)Cipollini, Gallo, and
  Palandri]{cipollini2020realized}
Fabrizio Cipollini, Giampiero~M Gallo, and Alessandro Palandri.
\newblock Realized variance modeling: Decoupling forecasting from estimation.
\newblock \emph{Journal of Financial Econometrics}, 18\penalty0 (3):\penalty0
  532--555, 2020.

\bibitem[Clements and Preve(2021)]{clements2021practical}
Adam Clements and Daniel~PA Preve.
\newblock A practical guide to harnessing the har volatility model.
\newblock \emph{Journal of Banking \& Finance}, 133:\penalty0 106285, 2021.

\bibitem[Corsi(2009)]{corsi2009simple}
Fulvio Corsi.
\newblock A simple approximate long-memory model of realized volatility.
\newblock \emph{Journal of Financial Econometrics}, 7\penalty0 (2):\penalty0
  174--196, 2009.

\bibitem[Dai et~al.(2018)Dai, Kozareva, Dai, Smola, and
  Song]{gnn_early_dai2018learning}
Hanjun Dai, Zornitsa Kozareva, Bo~Dai, Alex Smola, and Le~Song.
\newblock Learning steady-states of iterative algorithms over graphs.
\newblock In \emph{International Conference on Machine Learning}, pages
  1106--1114. PMLR, 2018.

\bibitem[Defferrard et~al.(2016)Defferrard, Bresson, and
  Vandergheynst]{gcn_defferrard2016GCN}
Micha{\"e}l Defferrard, Xavier Bresson, and Pierre Vandergheynst.
\newblock Convolutional neural networks on graphs with fast localized spectral
  filtering.
\newblock In \emph{Advances in Neural Information Processing Systems}, 2016.

\bibitem[Degiannakis and Filis(2017)]{degiannakis2017forecasting}
Stavros Degiannakis and George Filis.
\newblock Forecasting oil price realized volatility using information channels
  from other asset classes.
\newblock \emph{Journal of International Money and Finance}, 76:\penalty0
  28--49, 2017.

\bibitem[Diebold and Mariano(1995)]{diebold1995comparing}
Francis~X Diebold and Roberto~S Mariano.
\newblock Comparing predictive accuracy.
\newblock \emph{Journal of Business \& Economic Statistics}, 13\penalty0
  (3):\penalty0 253--263, 1995.

\bibitem[Engle and Kroner(1995)]{engle1995multivariate}
Robert~F Engle and Kenneth~F Kroner.
\newblock Multivariate simultaneous generalized arch.
\newblock \emph{Econometric Theory}, 11\penalty0 (1):\penalty0 122--150, 1995.

\bibitem[Fan et~al.(2014)Fan, Qi, and Xiu]{fan2014quasi}
Jianqing Fan, Lei Qi, and Dacheng Xiu.
\newblock Quasi-maximum likelihood estimation of {GARCH} models with
  heavy-tailed likelihoods.
\newblock \emph{Journal of Business \& Economic Statistics}, 32\penalty0
  (2):\penalty0 178--191, 2014.

\bibitem[Feng et~al.(2022)Feng, Chen, Li, Sarkar, and
  Zhang]{gnn_hop_feng2022powerful}
Jiarui Feng, Yixin Chen, Fuhai Li, Anindya Sarkar, and Muhan Zhang.
\newblock How powerful are {K}-hop message passing graph neural networks.
\newblock In \emph{Advances in Neural Information Processing Systems}, 2022.

\bibitem[Friedman et~al.(2008)Friedman, Hastie, and
  Tibshirani]{friedman2008sparse}
Jerome Friedman, Trevor Hastie, and Robert Tibshirani.
\newblock Sparse inverse covariance estimation with the graphical lasso.
\newblock \emph{Biostatistics}, 9\penalty0 (3):\penalty0 432--441, 2008.

\bibitem[Gu et~al.(2020)Gu, Kelly, and Xiu]{gu2020empirical}
Shihao Gu, Bryan Kelly, and Dacheng Xiu.
\newblock Empirical asset pricing via machine learning.
\newblock \emph{Review of Financial Studies}, 33\penalty0 (5):\penalty0
  2223--2273, 2020.

\bibitem[Hall and Yao(2003)]{hall2003inference}
Peter Hall and Qiwei Yao.
\newblock Inference in {ARCH} and {GARCH} models with heavy-tailed errors.
\newblock \emph{Econometrica}, 71\penalty0 (1):\penalty0 285--317, 2003.

\bibitem[Hansen and Dumitrescu(2022)]{hansen2022should}
Peter~Reinhard Hansen and Elena-Ivona Dumitrescu.
\newblock How should parameter estimation be tailored to the objective?
\newblock \emph{Journal of Econometrics}, 230\penalty0 (2):\penalty0 535--558,
  2022.

\bibitem[Hansen et~al.(2003)Hansen, Lunde, and Nason]{hansen2003choosing}
Peter~Reinhard Hansen, Asger Lunde, and James~M Nason.
\newblock Choosing the best volatility models: the model confidence set
  approach.
\newblock \emph{Oxford Bulletin of Economics and Statistics}, 65:\penalty0
  839--861, 2003.

\bibitem[Hansen et~al.(2011)Hansen, Lunde, and Nason]{hansen2011model}
Peter~Reinhard Hansen, Asger Lunde, and James~M Nason.
\newblock The model confidence set.
\newblock \emph{Econometrica}, 79\penalty0 (2):\penalty0 453--497, 2011.

\bibitem[Harvey et~al.(1997)Harvey, Leybourne, and Newbold]{harvey1997testing}
David Harvey, Stephen Leybourne, and Paul Newbold.
\newblock Testing the equality of prediction mean squared errors.
\newblock \emph{International Journal of Forecasting}, 13\penalty0
  (2):\penalty0 281--291, 1997.

\bibitem[Herskovic et~al.(2020)Herskovic, Kelly, Lustig, and
  Van~Nieuwerburgh]{herskovic2020firm}
Bernard Herskovic, Bryan Kelly, Hanno Lustig, and Stijn Van~Nieuwerburgh.
\newblock Firm volatility in granular networks.
\newblock \emph{Journal of Political Economy}, 128\penalty0 (11):\penalty0
  4097--4162, 2020.

\bibitem[Kingma and Ba(2014)]{gnn_train_adam_kingma2014adam}
Diederik~P Kingma and Jimmy Ba.
\newblock Adam: A method for stochastic optimization.
\newblock \emph{arXiv preprint arXiv:1412.6980}, 2014.

\bibitem[Kipf and Welling(2017)]{gcn_Kipf2017GCN}
Thomas~N. Kipf and Max Welling.
\newblock {Semi-Supervised Classification with Graph Convolutional Networks}.
\newblock In \emph{International Conference on Learning Representations}, 2017.

\bibitem[Li et~al.(2018)Li, Han, and Wu]{gnn_smoothness_li2018deeper}
Qimai Li, Zhichao Han, and Xiao-Ming Wu.
\newblock Deeper insights into graph convolutional networks for semi-supervised
  learning.
\newblock In \emph{AAAI Conference on Artificial Intelligence}, 2018.

\bibitem[Li and Tang(2021)]{li2021forecasting}
Sophia~Zhengzi Li and Yushan Tang.
\newblock Automated volatility forecasting.
\newblock \emph{Available at SSRN 3776915}, 2021.

\bibitem[Liang et~al.(2021)Liang, Zeng, Zhong, Chi, Feng, Ao, and
  Tang]{gnn_finance_risk_liang2021credit}
Ting Liang, Guanxiong Zeng, Qiwei Zhong, Jianfeng Chi, Jinghua Feng, Xiang Ao,
  and Jiayu Tang.
\newblock Credit risk and limits forecasting in e-commerce consumer lending
  service via multi-view-aware mixture-of-experts nets.
\newblock In \emph{Proceedings of the 14th ACM International Conference on Web
  Search and Data Mining}, pages 229--237, 2021.

\bibitem[Ling and McAleer(2003)]{ling2003asymptotic}
Shiqing Ling and Michael McAleer.
\newblock {Asymptotic theory for a vector ARMA-GARCH model}.
\newblock \emph{Econometric Theory}, 19\penalty0 (2):\penalty0 280--310, 2003.

\bibitem[Liu et~al.(2015)Liu, Patton, and Sheppard]{liu2015does}
Lily~Y Liu, Andrew~J Patton, and Kevin Sheppard.
\newblock Does anything beat 5-minute {RV}? {A} comparison of realized measures
  across multiple asset classes.
\newblock \emph{Journal of Econometrics}, 187\penalty0 (1):\penalty0 293--311,
  2015.

\bibitem[Liu et~al.(2018)Liu, Chen, Yang, Zhou, Li, and
  Song]{gnn_fraud_liu2018heterogeneous}
Ziqi Liu, Chaochao Chen, Xinxing Yang, Jun Zhou, Xiaolong Li, and Le~Song.
\newblock Heterogeneous graph neural networks for malicious account detection.
\newblock In \emph{Proceedings of the 27th ACM International Conference on
  Information and Knowledge Management}, pages 2077--2085, 2018.

\bibitem[Liu et~al.(2019)Liu, Chen, Li, Zhou, Li, Song, and
  Qi]{gnn_fraud_liu2019geniepath}
Ziqi Liu, Chaochao Chen, Longfei Li, Jun Zhou, Xiaolong Li, Le~Song, and Yuan
  Qi.
\newblock Geniepath: Graph neural networks with adaptive receptive paths.
\newblock In \emph{Proceedings of the AAAI Conference on Artificial
  Intelligence}, pages 4424--4431, 2019.

\bibitem[Masters and Luschi(2018)]{gnn_train_batch_masters2018revisiting}
Dominic Masters and Carlo Luschi.
\newblock Revisiting small batch training for deep neural networks.
\newblock \emph{arXiv preprint arXiv:1804.07612}, 2018.

\bibitem[Pascalau and Poirier(2021)]{pascalau2021increasing}
Razvan Pascalau and Ryan Poirier.
\newblock Increasing the information content of realized volatility forecasts.
\newblock \emph{Journal of Financial Econometrics}, 2021.

\bibitem[Patton(2011)]{patton2011volatility}
Andrew~J Patton.
\newblock Volatility forecast comparison using imperfect volatility proxies.
\newblock \emph{Journal of Econometrics}, 160\penalty0 (1):\penalty0 246--256,
  2011.

\bibitem[Patton and Sheppard(2009)]{patton2009evaluating}
Andrew~J Patton and Kevin Sheppard.
\newblock Evaluating volatility and correlation forecasts.
\newblock In \emph{Handbook of Financial Time Series}, pages 801--838.
  Springer, 2009.

\bibitem[Sawhney et~al.(2020)Sawhney, Agarwal, Wadhwa, and
  Shah]{gnn_stock_sawhney2020deep}
Ramit Sawhney, Shivam Agarwal, Arnav Wadhwa, and Rajiv Shah.
\newblock Deep attentive learning for stock movement prediction from social
  media text and company correlations.
\newblock In \emph{Proceedings of the 2020 Conference on Empirical Methods in
  Natural Language Processing (EMNLP)}, pages 8415--8426, 2020.

\bibitem[Scarselli et~al.(2008)Scarselli, Gori, Tsoi, Hagenbuchner, and
  Monfardini]{gnn_early_scarselli2008graph}
Franco Scarselli, Marco Gori, Ah~Chung Tsoi, Markus Hagenbuchner, and Gabriele
  Monfardini.
\newblock The graph neural network model.
\newblock \emph{IEEE Transactions on Neural Networks}, 20\penalty0
  (1):\penalty0 61--80, 2008.

\bibitem[Sheppard(2010)]{sheppard2010financial}
Kevin Sheppard.
\newblock Financial econometrics notes.
\newblock \emph{University of Oxford}, pages 333--426, 2010.

\bibitem[Shuman et~al.(2013)Shuman, Narang, Frossard, Ortega, and
  Vandergheynst]{gnn_survey_shuman2013emerging}
David~I Shuman, Sunil~K Narang, Pascal Frossard, Antonio Ortega, and Pierre
  Vandergheynst.
\newblock The emerging field of signal processing on graphs: Extending
  high-dimensional data analysis to networks and other irregular domains.
\newblock \emph{IEEE Signal Processing Magazine}, 30\penalty0 (3):\penalty0
  83--98, 2013.

\bibitem[Symitsi et~al.(2018)Symitsi, Symeonidis, Kourtis, and
  Markellos]{benchmark2018Symitsi}
Efthymia Symitsi, Lazaros Symeonidis, Apostolos Kourtis, and Raphael Markellos.
\newblock Covariance forecasting in equity markets.
\newblock \emph{Journal of Banking \& Finance}, 96:\penalty0 153--168, 2018.

\bibitem[Varneskov and Voev(2013)]{varneskov2013role}
Rasmus Varneskov and Valeri Voev.
\newblock The role of realized ex-post covariance measures and dynamic model
  choice on the quality of covariance forecasts.
\newblock \emph{Journal of Empirical Finance}, 20:\penalty0 83--95, 2013.

\bibitem[Wang et~al.(2019)Wang, Lin, Cui, Jia, Wang, Fang, Yu, Zhou, Yang, and
  Qi]{gnn_finance_risk_wang2019semi}
Daixin Wang, Jianbin Lin, Peng Cui, Quanhui Jia, Zhen Wang, Yanming Fang, Quan
  Yu, Jun Zhou, Shuang Yang, and Yuan Qi.
\newblock A semi-supervised graph attentive network for financial fraud
  detection.
\newblock In \emph{2019 IEEE International Conference on Data Mining (ICDM)},
  pages 598--607. IEEE, 2019.

\bibitem[Wang et~al.(2018)Wang, Wei, Wu, and Yin]{wang2018oil}
Yudong Wang, Yu~Wei, Chongfeng Wu, and Libo Yin.
\newblock Oil and the short-term predictability of stock return volatility.
\newblock \emph{Journal of Empirical Finance}, 47:\penalty0 90--104, 2018.

\bibitem[Wilms et~al.(2021)Wilms, Rombouts, and Croux]{wilms2021multivariate}
Ines Wilms, Jeroen Rombouts, and Christophe Croux.
\newblock Multivariate volatility forecasts for stock market indices.
\newblock \emph{International Journal of Forecasting}, 37\penalty0
  (2):\penalty0 484--499, 2021.

\bibitem[Wu et~al.(2020)Wu, Pan, Chen, Long, Zhang, and
  Philip]{gnn_wu2020survey}
Zonghan Wu, Shirui Pan, Fengwen Chen, Guodong Long, Chengqi Zhang, and S~Yu
  Philip.
\newblock A comprehensive survey on graph neural networks.
\newblock \emph{IEEE Transactions on Neural Networks and Learning Systems},
  32\penalty0 (1):\penalty0 4--24, 2020.

\bibitem[Zhang et~al.(2022)Zhang, Pu, Cucuringu, and Dong]{GHAR_zhang2022graph}
Chao Zhang, Xingyue~Stacy Pu, Mihai Cucuringu, and Xiaowen Dong.
\newblock Graph-based methods for forecasting realized covariances.
\newblock \emph{Available at SSRN}, 2022.

\bibitem[Zhang et~al.(2023)Zhang, Zhang, Cucuringu, and
  Qian]{zhang2022volatility}
Chao Zhang, Yihuang Zhang, Mihai Cucuringu, and Zhongmin Qian.
\newblock Volatility forecasting with machine learning and intraday
  commonality.
\newblock \emph{Journal of Financial Econometrics, forthcoming}, 2023.

\end{thebibliography}

\appendix
\renewcommand\thefigure{\thesection.\arabic{figure}}    
\renewcommand\thetable{\thesection.\arabic{table}}    

%%%%%%%%%%%%%%%%%%%%%%%%%%%%%%%%%%%%%%%%%%%%%%%%%%%%%%%%%%%%%%%%%%%%%%%%%%%%%%%%%%%%%%%%%
%%%%%%%%%%%%%%%%%%%%%%%%%%% GIN %%%%%%%%%%%%%%%%%%%%%%%%%%%%%%%%%%%%%%%
%%%%%%%%%%%%%%%%%%%%%%%%%%%%%%%%%%%%%%%%%%%%%%%%%%%%%%%%%%%%%%%%%%%%%%%%%%%%%%%%%%%%%%%%%
% \setcounter{figure}{0}    
% \setcounter{table}{0}
% % GIN
% \paragraph{Graph Isomorphism Network} (GIN), proposed by 
% \cite{gnn_gin_xu2018powerful}, assumes different weights in propagating information from neighbor nodes and self-connections, simply by adding a trainable scalar $\epsilon^{(l)}$ such that
% % in order to address the problem that many GNNs variants, including GCN,  have limited expressive power in distinguishing simple graph structures. In particular, GIN
% \begin{equation}
% \label{eq:gnn_literature_gin}
% \vspace{-2mm}
%     \boldsymbol{H}^{
% (l+1)} = \sigma \Big( \big( (1 + \epsilon^{(l)})\boldsymbol{H}^{(l)} + \boldsymbol{AH}^{(l)}\big) \boldsymbol{\Theta}^{(l)} \Big),
% \vspace{-2mm}
% \end{equation}%
% where $\boldsymbol{A}$ is the adjacency matrix without self-connections, and $\boldsymbol{\Theta}^{(l)} \in \mathbb{R}^{D^{(l)} \times D^{(l+1)}}$ is a layer-specific trainable parameter. 

%%%%%%%%%%%%%%%%%%%%%%%%%%%%%%%%%%%%%%%%%%%%%%%%%%%%%%%%%%%%%%%%%%%%%%%%%%%%%%%%%%%%%%%%%
%%%%%%%%%%%%%%%%%%%%%%%%%%% Data Statistics %%%%%%%%%%%%%%%%%%%%%%%%%%%%%%%%%%%%%%%
%%%%%%%%%%%%%%%%%%%%%%%%%%%%%%%%%%%%%%%%%%%%%%%%%%%%%%%%%%%%%%%%%%%%%%%%%%%%%%%%%%%%%%%%%
\setcounter{figure}{0}    
\setcounter{table}{0}
\section{Data statistics}\label{sec:data_stats}

\begin{landscape}
\begin{table}[H]
    \centering
    \caption{Summary statistics of realized volatility.}
    \resizebox{1.1\textwidth}{!}{% Please add the following required packages to your document preamble:
\begin{tabular}{ c||c }   % top level tables, with 2 columns  
% leftmost table of the top level table
\begin{tabular}{@{}cccccccccc@{}}
\toprule
Ticker & Mean & Std   & Min  & 25\% & 50\% & 75\% & Max    & DJIA                      & S\&P100     \\ \midrule
AAPL   & 2.30 & 3.39  & 0.07 & 0.70 & 1.25 & 2.46 & 38.30  & \checkmark & \checkmark \\
ABT    & 1.41 & 1.95  & 0.12 & 0.57 & 0.89 & 1.50 & 34.32  &                           & \checkmark \\
ACN    & 1.72 & 2.79  & 0.14 & 0.58 & 0.92 & 1.76 & 54.88  &                           & \checkmark \\
ADBE   & 2.53 & 3.34  & 0.16 & 0.93 & 1.54 & 2.76 & 45.55  &                           & \checkmark \\
ADP    & 1.41 & 2.51  & 0.10 & 0.49 & 0.78 & 1.39 & 44.36  &                           & \checkmark \\
AMGN   & 1.91 & 2.34  & 0.16 & 0.82 & 1.27 & 2.14 & 33.44  & \checkmark & \checkmark \\
AMT    & 2.16 & 3.83  & 0.19 & 0.68 & 1.11 & 2.10 & 53.19  &                           & \checkmark \\
AMZN   & 3.22 & 4.48  & 0.11 & 1.02 & 1.84 & 3.59 & 62.14  &                           & \checkmark \\
AXP    & 3.19 & 6.32  & 0.12 & 0.64 & 1.15 & 2.67 & 91.45  & \checkmark & \checkmark \\
BA     & 2.69 & 5.00  & 0.13 & 0.78 & 1.35 & 2.60 & 90.65  & \checkmark & \checkmark \\
BAC    & 4.93 & 11.48 & 0.10 & 1.01 & 1.81 & 3.68 & 135.30 &                           & \checkmark \\
BDX    & 1.37 & 1.84  & 0.13 & 0.54 & 0.86 & 1.48 & 28.52  &                           & \checkmark \\
BMY    & 1.77 & 2.20  & 0.08 & 0.72 & 1.14 & 1.93 & 30.75  &                           & \checkmark \\
BSX    & 3.15 & 4.39  & 0.20 & 1.14 & 1.92 & 3.35 & 55.28  &                           & \checkmark \\
C      & 5.48 & 14.6  & 0.15 & 0.99 & 1.82 & 3.94 & 257.34 &                           & \checkmark \\
CAT    & 2.79 & 4.00  & 0.15 & 0.94 & 1.58 & 2.89 & 45.26  & \checkmark & \checkmark \\
CB     & 1.82 & 3.66  & 0.07 & 0.44 & 0.75 & 1.62 & 61.54  &                           & \checkmark \\
CI     & 3.65 & 6.92  & 0.19 & 1.01 & 1.75 & 3.28 & 164.21 &                           & \checkmark \\
CMCSA  & 2.35 & 3.57  & 0.13 & 0.78 & 1.29 & 2.47 & 43.26  &                           & \checkmark \\
CME    & 3.07 & 5.49  & 0.18 & 0.84 & 1.38 & 2.72 & 68.79  &                           & \checkmark \\
COP    & 3.12 & 5.18  & 0.16 & 0.98 & 1.71 & 3.26 & 75.84  &                           & \checkmark \\
COST   & 1.44 & 2.11  & 0.0  & 0.51 & 0.79 & 1.44 & 26.30  &                           & \checkmark \\
CRM    & 4.00 & 4.93  & 0.22 & 1.44 & 2.41 & 4.64 & 61.67  & \checkmark & \checkmark \\
CSCO   & 1.98 & 2.92  & 0.14 & 0.70 & 1.13 & 2.09 & 43.74  & \checkmark & \checkmark \\
CVS    & 1.99 & 3.15  & 0.13 & 0.70 & 1.17 & 2.03 & 53.28  &                           & \checkmark \\
CVX    & 2.03 & 3.51  & 0.13 & 0.61 & 1.07 & 2.04 & 48.07  & \checkmark & \checkmark \\
D      & 1.44 & 2.56  & 0.1  & 0.56 & 0.85 & 1.40 & 40.39  &                           & \checkmark \\
DHR    & 1.6  & 2.41  & 0.14 & 0.54 & 0.95 & 1.67 & 29.78  &                           & \checkmark \\
DIS    & 1.89 & 3.04  & 0.12 & 0.60 & 1.01 & 1.88 & 40.56  & \checkmark & \checkmark \\
DUK    & 1.32 & 2.20  & 0.06 & 0.50 & 0.78 & 1.32 & 36.07  &                           & \checkmark \\
FIS    & 1.89 & 3.48  & 0.15 & 0.59 & 0.97 & 1.74 & 62.40  &                           & \checkmark \\
FISV   & 1.71 & 2.82  & 0.15 & 0.58 & 0.93 & 1.69 & 53.36  &                           & \checkmark \\
GE     & 3.08 & 5.54  & 0.09 & 0.68 & 1.43 & 3.05 & 77.33  &                           & \checkmark \\
GILD   & 2.36 & 2.67  & 0.23 & 1.03 & 1.55 & 2.64 & 33.62  &                           & \checkmark \\
GOOG   & 1.94 & 2.72  & 0.11 & 0.64 & 1.08 & 2.07 & 30.36  &                           & \checkmark \\
GS     & 3.24 & 6.27  & 0.19 & 0.92 & 1.49 & 2.81 & 112.41 & \checkmark & \checkmark \\
HD     & 2.11 & 3.59  & 0.15 & 0.62 & 1.02 & 2.01 & 48.22  & \checkmark & \checkmark \\
HON    & 1.85 & 3.25  & 0.1  & 0.52 & 0.97 & 1.84 & 49.64  & \checkmark & \checkmark \\
IBM    & 1.38 & 2.33  & 0.11 & 0.47 & 0.75 & 1.34 & 30.22  & \checkmark & \checkmark \\
INTC   & 2.29 & 3.12  & 0.14 & 0.86 & 1.39 & 2.44 & 42.90  & \checkmark & \checkmark \\
INTU   & 2.00 & 2.81  & 0.15 & 0.75 & 1.22 & 2.15 & 38.91  &                           & \checkmark \\ \bottomrule
\end{tabular} &  % starting rightmost sub table
% table 2
\begin{tabular}{@{}cccccccccc@{}}
\toprule
Ticker & Mean & Std   & Min  & 25\% & 50\% & 75\% & Max    & DJIA                      & S\&P100     \\ \midrule
ISRG   & 3.19 & 4.31  & 0.22 & 1.10 & 1.81 & 3.38 & 46.66  &                           & \checkmark \\
JNJ    & 0.92 & 1.56  & 0.06 & 0.35 & 0.54 & 0.90 & 24.74  & \checkmark & \checkmark \\
JPM    & 3.46 & 7.04  & 0.15 & 0.74 & 1.36 & 2.82 & 108.17 & \checkmark & \checkmark \\
KO     & 0.99 & 1.68  & 0.07 & 0.37 & 0.58 & 1.00 & 25.00  & \checkmark & \checkmark \\
LLY    & 1.59 & 2.29  & 0.13 & 0.61 & 0.98 & 1.70 & 35.90  &                           & \checkmark \\
LMT    & 1.64 & 2.59  & 0.12 & 0.56 & 0.94 & 1.64 & 35.79  &                           & \checkmark \\
LOW    & 2.71 & 4.20  & 0.17 & 0.88 & 1.45 & 2.77 & 73.32  &                           & \checkmark \\
MA     & 2.86 & 4.60  & 0.13 & 0.73 & 1.31 & 2.79 & 52.20  &                           & \checkmark \\
MCD    & 1.17 & 2.15  & 0.08 & 0.39 & 0.61 & 1.13 & 37.57  & \checkmark & \checkmark \\
MDT    & 1.5  & 2.19  & 0.13 & 0.59 & 0.93 & 1.57 & 36.66  &                           & \checkmark \\
MMM    & 1.43 & 2.25  & 0.08 & 0.46 & 0.81 & 1.49 & 31.11  & \checkmark & \checkmark \\
MO     & 1.41 & 2.25  & 0.06 & 0.52 & 0.84 & 1.43 & 39.67  &                           & \checkmark \\
MRK    & 1.65 & 2.45  & 0.12 & 0.58 & 0.92 & 1.74 & 30.99  & \checkmark & \checkmark \\
MS     & 5.74 & 14.30 & 0.20 & 1.25 & 2.18 & 4.33 & 286.91 &                           & \checkmark \\
MSFT   & 1.82 & 2.51  & 0.11 & 0.67 & 1.09 & 1.92 & 30.64  & \checkmark & \checkmark \\
NFLX   & 5.53 & 5.69  & 0.36 & 2.14 & 3.78 & 6.75 & 72.86  &                           & \checkmark \\
NKE    & 2.03 & 3.00  & 0.14 & 0.74 & 1.15 & 2.02 & 47.87  & \checkmark & \checkmark \\
NVDA   & 5.14 & 6.03  & 0.40 & 1.83 & 3.2  & 5.96 & 72.25  &                           & \checkmark \\
ORCL   & 1.90 & 2.84  & 0.08 & 0.66 & 1.14 & 2.05 & 44.23  &                           & \checkmark \\
PEP    & 1.02 & 1.78  & 0.06 & 0.37 & 0.58 & 1.01 & 28.18  &                           & \checkmark \\
PFE    & 1.55 & 2.07  & 0.14 & 0.59 & 0.95 & 1.67 & 26.54  &                           & \checkmark \\
PG     & 1.00 & 1.76  & 0.09 & 0.38 & 0.58 & 0.98 & 31.60  & \checkmark & \checkmark \\
PNC    & 3.64 & 7.52  & 0.16 & 0.79 & 1.38 & 3.04 & 141.27 &                           & \checkmark \\
QCOM   & 2.46 & 3.39  & 0.10 & 0.81 & 1.49 & 2.77 & 42.15  &                           & \checkmark \\
SBUX   & 2.45 & 3.90  & 0.18 & 0.71 & 1.24 & 2.48 & 63.45  &                           & \checkmark \\
SO     & 1.19 & 1.98  & 0.12 & 0.47 & 0.72 & 1.22 & 36.40  &                           & \checkmark \\
SYK    & 1.67 & 2.61  & 0.08 & 0.62 & 0.98 & 1.76 & 49.51  &                           & \checkmark \\
T      & 1.49 & 2.55  & 0.08 & 0.47 & 0.76 & 1.39 & 32.03  &                           & \checkmark \\
TGT    & 2.46 & 4.02  & 0.11 & 0.76 & 1.24 & 2.34 & 53.02  &                           & \checkmark \\
TJX    & 2.33 & 3.34  & 0.16 & 0.76 & 1.24 & 2.53 & 55.49  &                           & \checkmark \\
TMO    & 1.89 & 2.74  & 0.16 & 0.71 & 1.14 & 1.99 & 40.82  &                           & \checkmark \\
TRV    & 2.04 & 4.09  & 0.11 & 0.49 & 0.81 & 1.76 & 57.95  & \checkmark &                           \\
TXN    & 2.33 & 3.02  & 0.16 & 0.84 & 1.41 & 2.57 & 48.68  &                           & \checkmark \\
UNH    & 2.70 & 4.34  & 0.16 & 0.78 & 1.35 & 2.57 & 52.54  & \checkmark & \checkmark \\
UNP    & 2.53 & 3.94  & 0.14 & 0.83 & 1.39 & 2.52 & 45.94  &                           & \checkmark \\
UPS    & 1.58 & 2.35  & 0.10 & 0.51 & 0.88 & 1.72 & 31.67  &                           & \checkmark \\
USB    & 3.20 & 6.88  & 0.13 & 0.62 & 1.16 & 2.64 & 95.38  &                           & \checkmark \\
VZ     & 1.40 & 2.36  & 0.10 & 0.50 & 0.77 & 1.33 & 34.19  & \checkmark & \checkmark \\
WFC    & 4.05 & 8.89  & 0.11 & 0.73 & 1.39 & 3.24 & 106.81 &                           & \checkmark \\
WMT    & 1.18 & 1.76  & 0.11 & 0.45 & 0.67 & 1.18 & 27.18  & \checkmark & \checkmark \\ 
    &  &   &  & & & & &  &  \\ \bottomrule
\end{tabular}\\
\end{tabular}}
    \caption*{\textit{Note:} The table reports summary statistics for the daily realized volatility of stocks in DJIA30 or S\&P100. The statistics are averaged across each trading day.}
    \label{tab:data_stats}
\end{table}
\end{landscape}

\begin{table}[H]
    \centering
    % \caption{Summary statistics on the distribution of the shortest path distance, across all pairs of nodes in the graph (as a fraction of all pairwise interactions). For example, in the case of S\&P100, 12\% of pairs of nodes have their shortest path distance of size 3.}
    \caption{Frequency (in percentage) of the shortest path distance.}
    \resizebox{0.4\textwidth}{!}{\begin{tabular}{lccccc}
\toprule
SPD & 1 & 2 & 3 & 4 & 5 \\
\midrule
DJIA & 57.7 & 41.8 & 0.5 & 0.0 & 0.0 \\
S\&P100 & 24.3 & 61.2 & 12.0 & 2.2 & 0.3  \\

\bottomrule
\end{tabular}
}
    \label{tab:spd_freq}
    \caption*{\textit{Note:} For example, in the case of S\&P100, 12\% of pairs of nodes have their shortest path distance of size 3.}
\end{table}

%%%%%%%%%%%%%%%%%%%%%%%%%%%%%%%%%%%%%%%%%%%%%%%%%%%%%%%%%%%%%%%%%%%%%%%%%%%%%%%%%%%%%%%%%
%%%%%%%%%%%%%%%%%%%%%%%%%%% Hyperparameter %%%%%%%%%%%%%%%%%%%%%%%%%%%%%%%%%%%%%%%
%%%%%%%%%%%%%%%%%%%%%%%%%%%%%%%%%%%%%%%%%%%%%%%%%%%%%%%%%%%%%%%%%%%%%%%%%%%%%%%%%%%%%%%%%
\setcounter{figure}{0}    
\setcounter{table}{0}
\section{Hyperparameter tuning} \label{sec:hyperparameter_GNN}

Following the convention of stochastic optimization (\cite{gnn_train_adam_kingma2014adam}), we set the batch size to 32.\footnote{Mini-batch training is believed to improve generalization performance, see \cite{gnn_train_batch_masters2018revisiting}.} The learning rate for Adam is set to be $10^{-3}$. We stop the training procedure early if there is a sign of overfitting, that is, the training loss keeps dropping but validation loss increases beyond a tolerance level. 

% To avoid overfitting, we apply grid search to find an optimal number of hidden neurons in the proposed GNNHARs. 

% \begin{figure}[H]
%     \centering
%     \caption{Validation performance under different dimensions of hidden representations in GNN} 
%     \subfigure[MSE\label{fig:nhid_mse}]{
%     \includegraphics[width=0.47\textwidth, trim=7mm 2mm 2mm 2mm,clip]{figures/nhid/nhid_mse.png}}
%     \subfigure[\label{fig:nhid_qlike}]{
%     \includegraphics[width=0.47\textwidth, trim=7mm 2mm 2mm 2mm,clip]{figures/nhid/nhid_qlike.png}}
%     \label{fig:nhid_losses}
%      \caption*{\textit{Note:} Each box is obtained from 10 replicated experiments with different random initial parameters. }
% \end{figure}

\begin{figure}[H]
    \centering
    \caption{Validation performance under different dimensions of hidden representations in $\text{GNNHAR1L}_M$.} 
    \includegraphics[width=0.47\textwidth]{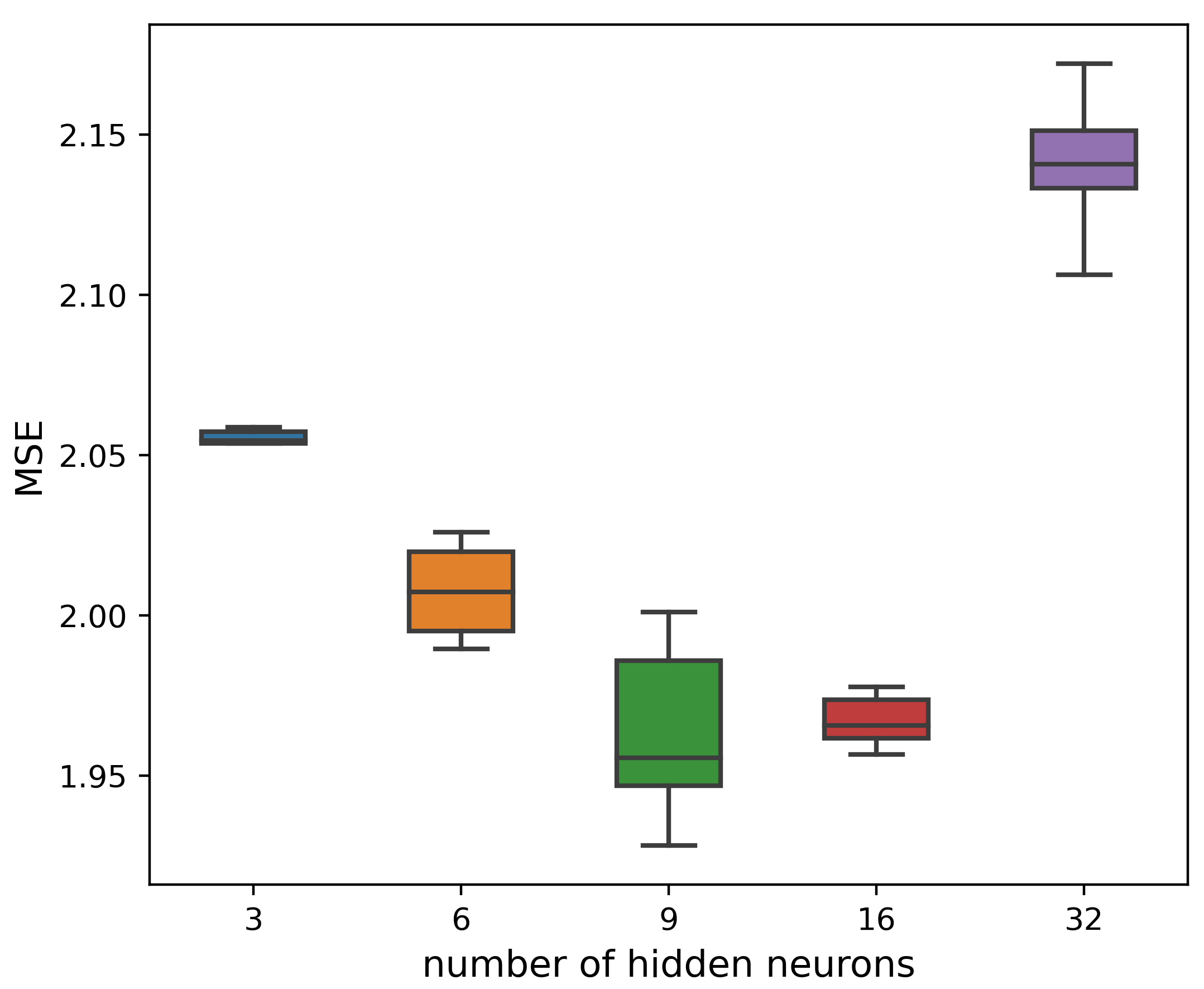}
    \includegraphics[width=0.47\textwidth]{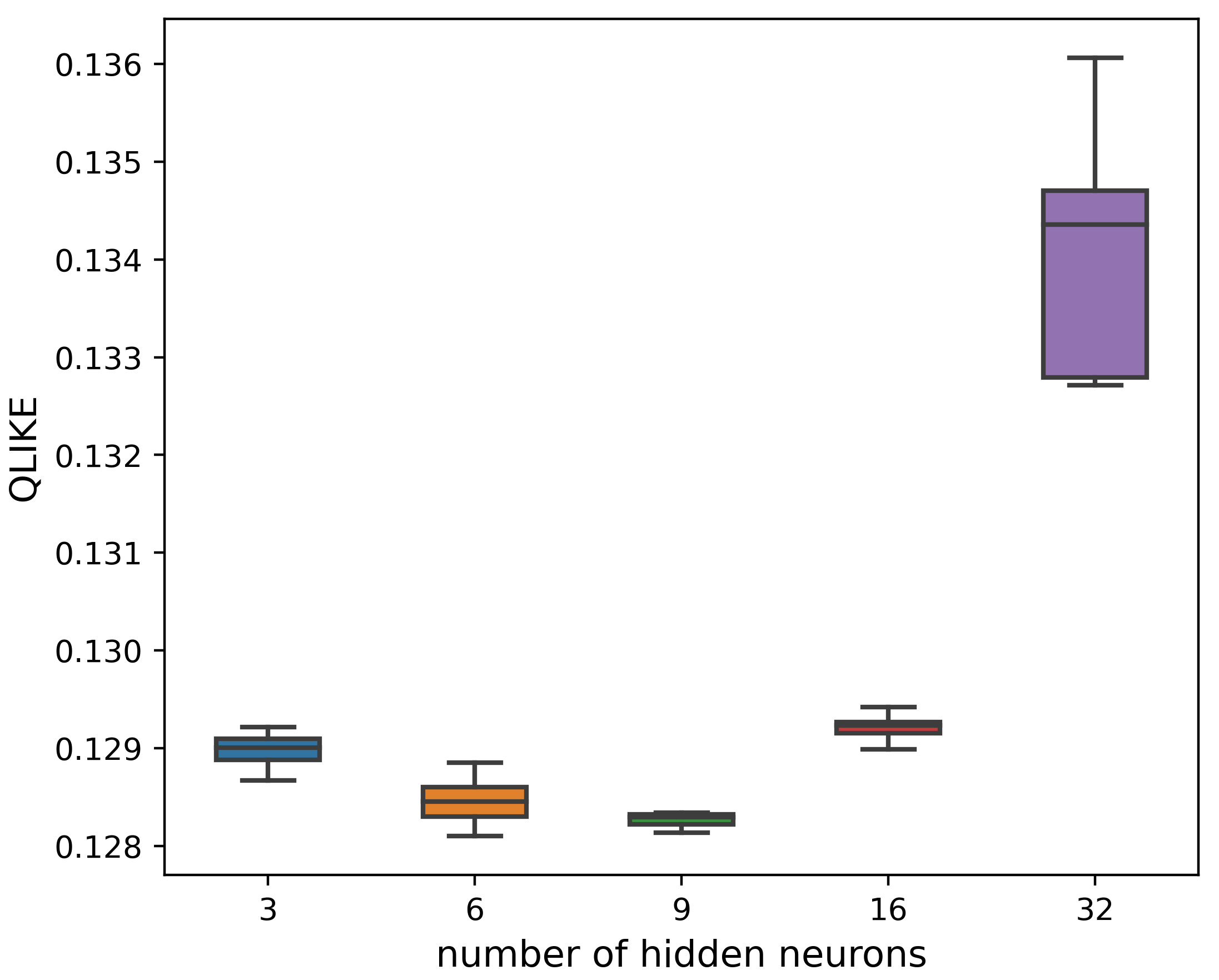}
    \label{fig:nhid_losses}
     \caption*{\textit{Note:} Each box is obtained from 10 replicated experiments with different random initial parameters. }
\end{figure}

To a large extent, the dimension of hidden representations or the number of hidden neurons in $l$-th layer, i.e. $D^{(l)}$ in \eqref{eq:GNN_layer} reflects the complexity of our models. Inadequate dimensions may lack the capability to effectively capture the underlying data structure, while excessively large dimensions could lead to overfitting and poor generalization performance. To mitigate this issue, we use a grid search over $D^{(l)} \in \{3, 6, 9, 16, 32\}$ on validation datasets. Figure \ref{fig:nhid_losses} shows that a hidden dimension of 9 in a one-layer GNNHAR model leads to the smallest MSE and QLIKE on the validation data. The same conclusion holds true for the QLIKE-trained models as well. When multiple GNN layers are utilized, we maintain the same $D^{(l)}$ value as determined in the one-layer model.

%%%%%%%%%%%%%%%%%%%%%%%%%%%%%%%%%%%%%%%%%%%%%%%%%%%%%%%%%%%%%%%%%%%%%%%%%%%%%%%%%%%%%%%%%
%%%%%%%%%%%%%%%%%%%%%%%%%%% FVU for longer horizons %%%%%%%%%%%%%%%%%%%%%%%%%%%%%%%%%%%%%%%
%%%%%%%%%%%%%%%%%%%%%%%%%%%%%%%%%%%%%%%%%%%%%%%%%%%%%%%%%%%%%%%%%%%%%%%%%%%%%%%%%%%%%%%%%
% \setcounter{figure}{0}    
% \setcounter{table}{0}
% \section{Nonlinearity over longer horizons} \label{sec:nonlin_longer}

% \begin{table}[H]
%     \centering
%     \caption{Stratified out-of-sample losses and FVU over longer horizons.}
%     \resizebox{1.0\textwidth}{!}{\input{tables_Jan/nonlinear_longer.tex}}
%     \caption*{\textit{Note:} The table reports stratified losses and the fraction of variance unexplained of multiple models (compared to HAR) for forecasting 1-week-ahead (Panel A) and 1-month-ahead (Panel B) realized volatilities across different market regimes.}
%     \label{tab:nonlin_longer}
% \end{table}

%%%%%%%%%%%%%%%%%%%%%%%%%%%%%%%%%%%%%%%%%%%%%%%%%%%%%%%%%%%%%%%%%%%%%%%%%%%%%%%%%%%%%%%%%
%%%%%%%%%%%%%%%%%%%%%%%%%%% GHAR2Hop %%%%%%%%%%%%%%%%%%%%%%%%%%%%%%%%%%%%%%%
%%%%%%%%%%%%%%%%%%%%%%%%%%%%%%%%%%%%%%%%%%%%%%%%%%%%%%%%%%%%%%%%%%%%%%%%%%%%%%%%%%%%%%%%%

\setcounter{figure}{0}    
\setcounter{table}{0}
\section{GHAR with multi-hop (GHAR2Hop)} \label{sec:ghar2hop}
It is important to highlight that HAR can be interpreted as a model that only considers the 0th-hop neighbors, i.e. the target node itself, while the GHAR takes into account both the 0th-hop and 1st-hop neighbors. In order to explore the potential benefits of multi-hop neighbors in enhancing volatility forecasting, we delve into the investigation of whether they provide additional predictive power. To address this novel question, we consider the following model.
\begin{equation}\label{eq:ghar_2hop}
\textbf{GHAR2Hop}(\boldsymbol{A}): \quad \boldsymbol{RV}_{t} = \boldsymbol{\alpha} + \boldsymbol{V}_{:t-1} \boldsymbol{\beta} + \boldsymbol{W} \boldsymbol{V}_{:t-1} \boldsymbol{\gamma} + \textsc{Hop2}(\boldsymbol{A}) \boldsymbol{V}_{:t-1} \boldsymbol{\delta} + \boldsymbol{u}_{t},
\end{equation}
where $\textsc{Hop2}(\boldsymbol{A})$ maps the raw adjacent matrix (for 1st-hop neighbors) to the adjacent matrix of 2nd-hop neighbors. Specifically,  $\textsc{Hop2}(\boldsymbol{A}) = \operatorname{XOR} (\boldsymbol{A}^2 \wedge (\neg \boldsymbol{A}), \boldsymbol{I}_N)$.  $\boldsymbol{A}^2[i, j]$ has a non-zero if it is possible to go from node $i$ to node $j$ in 2 or fewer steps, $\neg \boldsymbol{A}$ excludes the 1st-hop neighbors, and $\operatorname{XOR}$ confirms the diagonal of 2nd-hop adjacent matrix to be zero. For a visual representation and further details, we refer the reader to Example \ref{ex:adj_hop} and Figure \ref{fig:illustration_graphs_adj}. In our experiments, we use the normalized adjacent matrix of 2nd-hop neighbors and estimate \eqref{eq:ghar_2hop} through OLS.

\begin{figure}[H]
    \centering
    \caption{DM test between GHAR2Hop and GHAR.} 
    \includegraphics[width=0.7\textwidth, trim=3cm 5mm 3cm 2cm,clip]{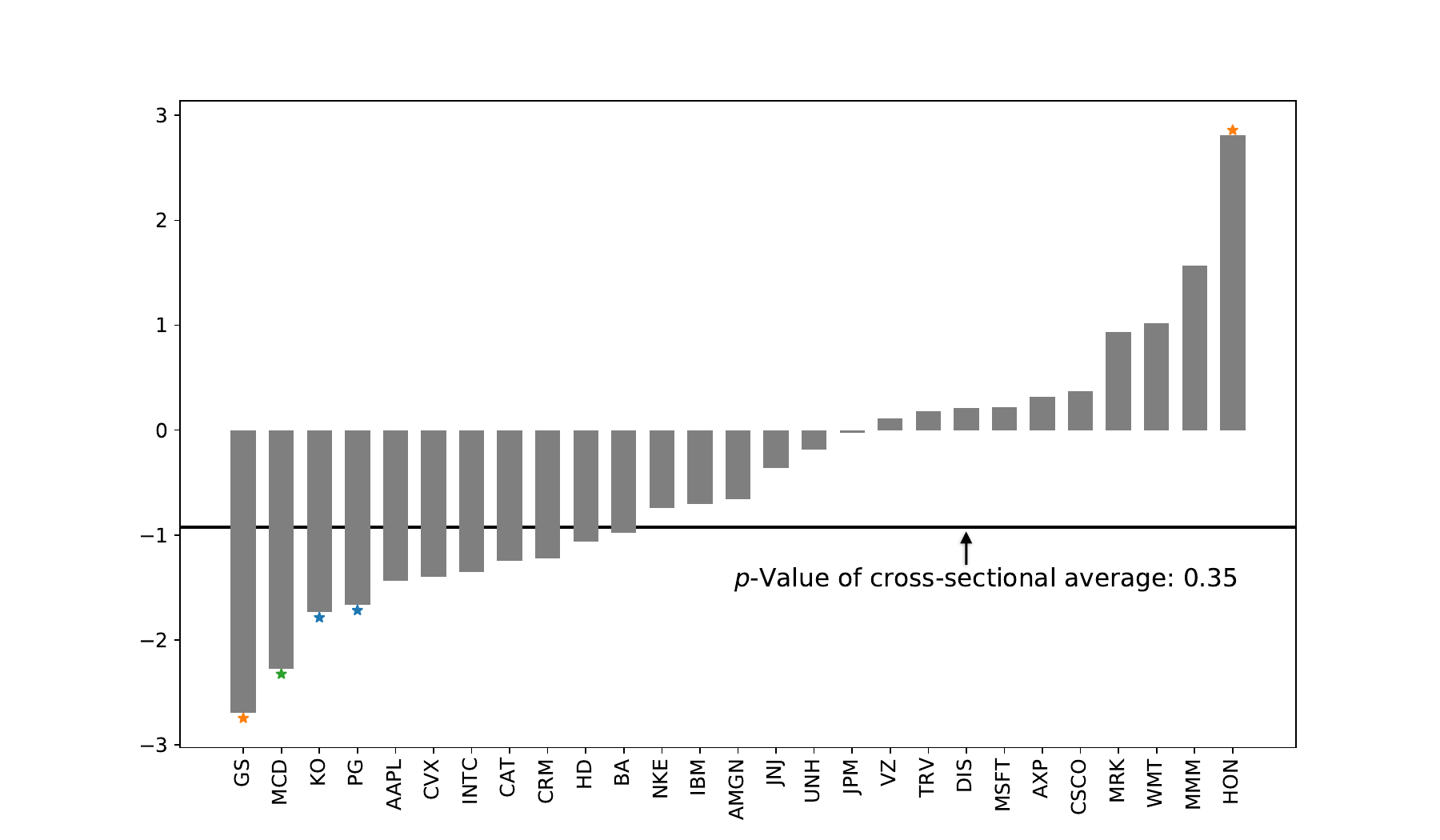}
    \caption*{\textit{Note:} A positive (negative) number indicates superiority for the GHAR (GHAR2Hop) model. The $y$-axis represents the DM test values based on QLIKEs between GHAR2Hop and GHAR, while the $x$-axis lists the stock symbols. Stars indicate the $p$-values, with orange, green, and blue representing significance at the 1\%, 5\%, and 10\% levels, respectively. The horizon line represents the cross-sectional DM test value and its corresponding $p$-value.}
    \label{fig:DM_2hop}
\end{figure}

The DM test results between GHAR2Hop and GHAR are presented in Figure \ref{fig:DM_2hop}. The cross-sectional DM test value is approximately -1, with a corresponding $p$-value of approximately 35\%. These results reinforce the primary findings regarding the role of multi-hop neighbors, indicating that including 2-hop neighbors may not provide substantial additional predictive power.

\begin{figure}[H]
    \centering
    \caption{Coefficients in GHAR2Hop.}
    \includegraphics[width=0.6\textwidth, trim=5mm 1cm 2cm 2cm,clip]{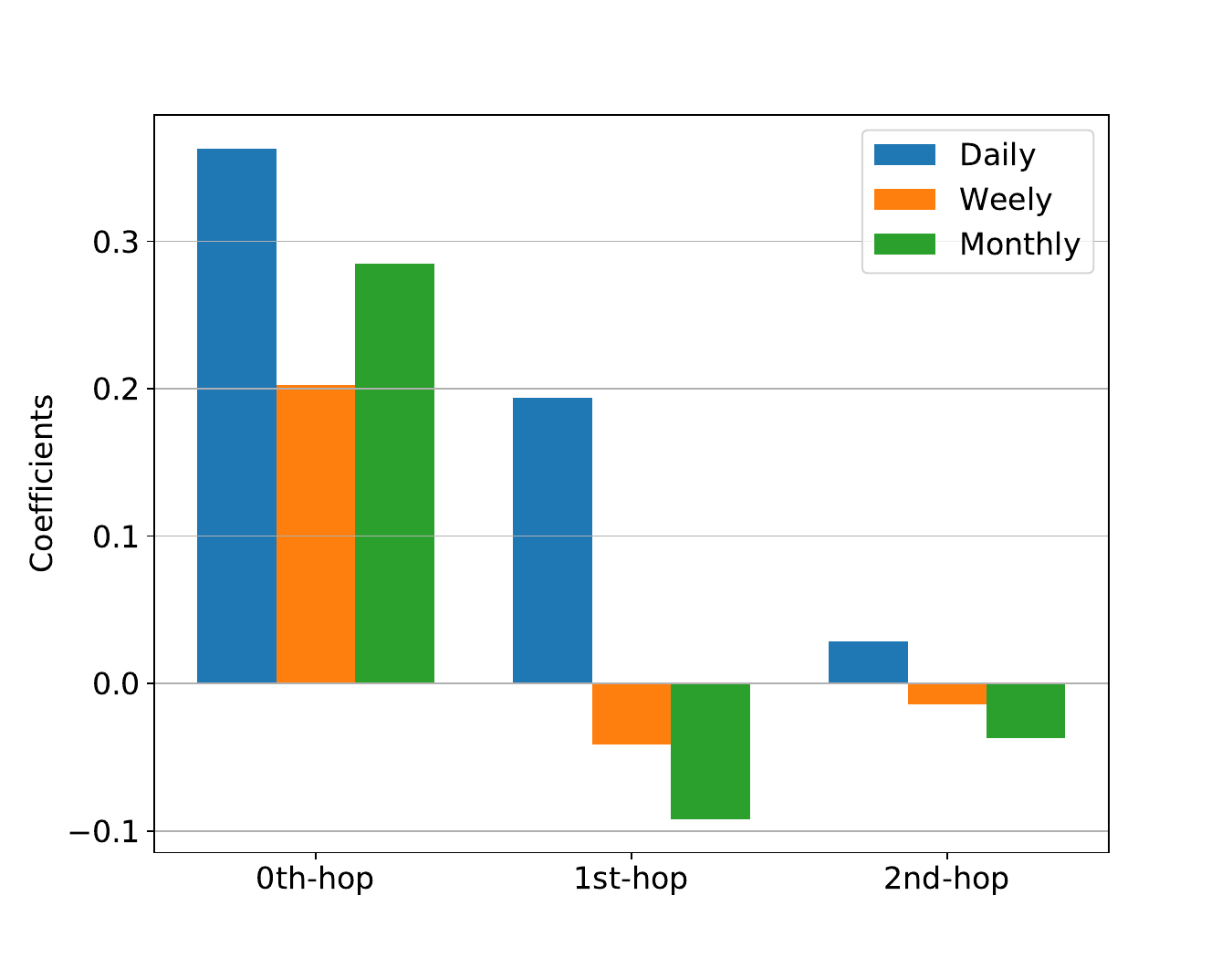}
    \caption*{\textit{Note:} This figure describes the average coefficients of different hop neighborhoods over multiple horizons.}
    \label{fig:coef_2hop}
\end{figure}

In Figure \ref{fig:coef_2hop}, we conduct a detailed examination of the coefficients associated with $K$-hop neighbors across different forecasting horizons. Based on the given definitions, the 0th-hop coefficients for the Daily (resp. Weekly, Monthly) horizon represent $\beta_d$ (resp. $\beta_w$, $\beta_m$), the 1st-hop coefficients correspond to $\gamma_d$ (resp. $\gamma_w$, $\gamma_m$), and the 2nd-hop coefficients denote $\delta_d$ (resp. $\delta_w$, $\delta_m$).
Figure \ref{fig:coef_2hop} reveals that the coefficients at 0th-hop are positive over three horizons (i.e. $\beta_d, \beta_w, \beta_m > 0$), consistent with previous literature (\cite{bollerslev2018modeling}). We also observe that the daily coefficients are positive on average but rapidly decay with distance (i.e. $\beta_d > \gamma_d > \delta_d$). Specifically, the daily coefficient associated with 2nd-hop neighbors is approximately 1/8 (1/16) relative to the coefficient of their 1st-hop (0th-hop) counterparts. Another interesting observation is that the weekly and monthly coefficients are negative, potentially due to high collinearity, as highlighted in \cite{GHAR_zhang2022graph}. Nonetheless, the magnitude of these coefficients diminishes as the distance increases, suggesting that the influence of the 2nd-hop neighbors may be negligible.

% \subsection{Rolling coefficients}
% \cite{GHAR_zhang2022graph} found that the model coefficients $\boldsymbol{\beta}$ in GHAR might capture the market conditions to some extent. Following their experimental settings,  in Figure \ref{fig:coef_nonlinear} we plot $\beta_d$, the coefficient for the past daily volatility, of GHAR and the proposed GNNHAR1L. Generally, the $\beta_d$ estimates obtained from the two models have a similar pattern from 2011 to 2021. 
% Three rapid shocks are detected at the end of 2012, 2015, and at early 2020. By comparing the blue and orange curves, it seems that GNNHAR1L is more responsive to market changes.  For example, $\beta_d$ of GNNHAR1L has dropped rapidly to a much lower level than its counterpart of GHAR from June to Oct 2015. Additionally, the dispersion in the coefficients of GNNHAR1L might correspond to the market sentiment. During turbulent periods, such as August 2015 and March 2020, the standard deviation of $\beta_d$ is higher than in other periods. It is during these turbulent periods that the nonlinearity effects come into play, and thus linear models begin to struggle with performance. 

% \begin{figure}[H]
%     \centering
%     \caption{Rolling coefficients of $\beta_d$.}
%     \includegraphics[width=0.75\textwidth]{figures/GHAR_GNNCoef_Beta.pdf}
%     \label{fig:coef_nonlinear}
%     \caption*{\textit{Note:} The blue (orange) curve represents the estimated $\beta_d$ of GHAR (GNNHAR), respectively. The barplot describes the standard deviation of $\beta_d$ computed from ensembling GNNHARs.} 
% \end{figure}

\end{document}